\documentclass[12pt, letter]{article}

\usepackage{geometry}
\usepackage{setspace}
\usepackage{stix}
\usepackage{titling}
\usepackage{authblk}
\usepackage{mathtools}
\usepackage{xcolor, hyperref}
\hypersetup{colorlinks=true, linkcolor=blue, citecolor=blue, urlcolor=cyan}
\usepackage{cite}
\usepackage{algpseudocode} 
\usepackage{algorithmicx}
\usepackage{algorithm}
\usepackage{graphicx}
\usepackage{subcaption}
\usepackage{siunitx}
\usepackage{indentfirst}
\usepackage{pgfgantt}
\usepackage[nottoc,numbib]{tocbibind}
\usepackage{enumitem}
\usepackage{cases}
\usepackage{multirow}

\DeclarePairedDelimiter{\ps}{\lparen}{\rparen}
\DeclarePairedDelimiter{\bs}{\lbrack}{\rbrack}

\DeclarePairedDelimiter{\abs}{\lvert}{\rvert} 
 
\DeclarePairedDelimiter{\br}{\{}{\}}

\makeatletter
\newcommand{\pushright}[1]{\ifmeasuring@#1\else\omit$\displaystyle#1$\ignorespaces\fi}
\makeatother

\title{Thermal Conductance Correlations of Static Lubricated Ball Bearings}

\author[1]{Indronil Ghosh\footnote{email: ighosh@g.ucla.edu}}
\author[2]{John P. McHale}
\author[2]{Yoshimi R. Takeuchi}
\author[2]{Peter P. Frantz}
\author[2]{Payton J. Batliner}
\author[1]{Timothy S. Fisher}
\affil[1]{University of California, Los Angeles}
\affil[2]{The Aerospace Corporation}

\date{ }
\begin{document}
\begin{titlepage}
\maketitle
\begin{abstract}
\noindent Ball bearings are commonly used to reduce the friction in rotating mechanical components. The present work reports improved numerical approaches to model bearing thermal conductance in the absence of convection. We start by modeling the thermal pathway across a single ball-to-race pathway for a simplified geometry, an azimuthally symmetric ball in contact with a flat surface (ball-on-flat). A first-principles approach is used to calculate the static lubricant meniscus shape using a custom-developed Python code. We apply the finite element method (FEM) to extract total thermal conductance of the lubricated ball-on-flat system. Using similar methods, we also present a three-dimensional numerical model of the lubricant meniscus in a static angular contact ball bearing section (ball-on-race). By generating thermal conductance correlations for Yovanovich's classic lubricated model, our two-dimensional multiphysics model, and our three-dimensional multiphysics model, we enable comparison of the results of all three models. To compare them, parametric studies are conducted to illustrate the effect of lubricant volume and applied load on the total thermal conductance for each model. The hierarchical methodology reported here improves both the fidelity of tribo-thermo-mechanical modeling and establishes a reference for the accuracy of commonly used geometric approximations for thermal transport in spacecraft ball bearings.

\end{abstract}
\end{titlepage}
\tableofcontents
\newpage
\newgeometry{top=1in,bottom=1in,right=1in,left=1in}

\section{Introduction}
The thermal constriction resistance between ball and bearing raceways produces poor heat removal rates, so that in most terrestrial applications, convection dominates the cooling process.  In spacecraft applications, where convective cooling is commonly unavailable, ball bearings provide the primary thermal conduction pathway between a shaft and housing. Developing a clear understanding of thermal conductance through a bearing is crucial for prediction of spacecraft bearing temperatures, which in turn affect life considerations for the mechanism.

Extensive theoretical research has been conducted on ball bearings, and more generally, on sliding contacts in both dry and wet configurations. Of historic significance is Yovanovich's \cite{Yovanovich1967, Yovanovich1970} analytical formulation of the thermal contact resistance between a dry, static ball and race through the use of Hertzian contact theory, dating back to 1967. Yovanovich \cite{MuzychkaYovanovich2001} also formulated an analytical solution of the thermal resistance of a dry, moving elliptic contact for any value of Peclet number. The temperatures within a dry, sliding contact have also been studied through application of Hertzian theory \cite{Francis1971, Bejan1989}, the finite element method \cite{Varadi1998}, finite difference techniques \cite{Salem1983}, Green's functions \cite{TianKennedy1994}, and other analytical and numerical approaches \cite{HarrisMindel1973, KannelBarber1989}. For the case of a static, lubricated ball-on-flat, Yovanovich \cite{YovanovichKitscha1973} proposed a thermal model for the thermal resistance of a lubricated, static ball on flat in vacuum or air, assuming the conduction pathway of the ball-flat contact and that of the lubricant to be independent of each other. The studies of lubricated, dynamic contacts fall under the discipline of thermoelastohydrodynamics and its topical subsets. While we do not consider moving surfaces in this study, numerous studies on the temperature effects on elastohydrodynamic lubrication, for varying loads, speeds, and boundary conditions are available \cite{Sternlicht1961, Snyder1965, Chang1965, DowsonWhitaker1965, Pascovici1974, SafarSzeri1974, MurchWilson1975, Kaludjercic1980, GhoshHamrock1985, SadeghiDow1987, SadeghiDowJohnson1987, WangZhang1987, SadeghiSui1990, PeiranShizhu1990}.

While Yovanovich and Kitscha's 1973 model \cite{YovanovichKitscha1973} of a lubricated ball-on-flat assumed a vertical oil to air or vacuum boundary, more recent work on liquid meniscus profiles and forces is available to represent the lubricant meniscus more accurately. Early literature on meniscus theory assumed spherical curvature \cite{BowdenTabor1950, BowdenThrossell1951, Deryagin1955, DerjaguinChuraev1976, FisherIsraelachvili1981, Christenson1985, Weisenhorn1989, Mate1989, Gee1990, MateNovotny1991, LiTalke1992, Woodward1992, Israelachvili1992, TianMatsudara1993, GuiMarchon1995, GaoTian1995, GaoBhushan1995, MatsuokaKato1996}, while the more recent theory of Gao et al. \cite{Gao1997, Gao1998} is novel in that it incorporates disjoining pressure into its meniscus force model, in addition to the Laplace pressure, thus elimating the spherical curvature assumption. We apply Gao's force model in this paper to calculate the meniscus geometry for the lubricant contacting the ball and flat in a ball-on-flat system.

Significant theoretical research has also occurred on the mechanical deformation of a sphere contacting a flat surface, commonly referred to as the elastic-plastic contact of a sphere and flat, or the problem of indenting a half-space with a sphere. The 1999 review by Liu et al. \cite{Liu1999} provides a survey of many models for simulating the mechanics of rough contact. Early works on this topic confined the analysis strictly to the elastic regime \cite{GreenwoodWilliamson1966, Yovanovich1967, TimoshenkoGoodier1970}, or the plastic regime \cite{AbbottFirestone1933, Pullen1972}. Later works aimed to address both regimes together and present unified analysis of the elastic-plastic contact, covering the intermediate ranges of deformation that are of broad interest \cite{Chang1987, Evseev1991, Kucharski1994, Chang1997, Zhao2000}. Kogut and Etsion's seminal work \cite{KogutEtsion2002} presents finite element method equations that accurately describe the elastic-plastic contact of a deformable sphere on a rigid flat as a function of geometric and material inputs. We opt to apply the Kogut-Etsion solution in this work to compute the load-dependent deformed ball geometry due to the model's applicability to both elastic and plastic deformation regimes, as well as its acceptance of a wide range of geometric and material parameters.

More recently, researchers have presented a Simulia Abaqus based thermal conductance analysis tool \cite{Bertagne2023}, allowing for prescription of a ball within races mechanical mesh, and static lubricant meniscus by extruding a lubricant cylinder and performing Boolean subtraction on it to remove the lubricant regions that would intersect the ball within races mesh. While this software is powerful for calculating the thermal conductance across whatever geometry a user specifies, we aim to fill important gaps of knowledge, namely: physics-based lubricant meniscus generation and convenient mathematical correlations of thermal conductance, for both two-dimensional (2D) azimuthally symmetric ball-on-flat geometries as well as three-dimensional (3D) ball-on-race geometries. With our methodology, we seek to empower thermal engineers and scientists with a technique for estimating thermal conductance in ball bearing geometries that is not only computationally efficient, but also predicts load-driven deformation and static lubricant meniscus formation with high fidelity of the underlying physics.

\section{Methods for ball-on-flat model} \label{Methods for ball-on-flat model}
The ball-on-flat multiphysics model consists of three main components: a mechanical deformation submodel, an azimuthally symmetric lubricant meniscus submodel, and a FEM solution to the steady state head conduction equation in cylindrical coordinates. In summary, the model executes the following steps:
\begin{enumerate}
    \item User inputs thermal conductivities, ball radius, lubricant volume.
    \item Obtain the deformed ball geometry, using the Kogut-Etsion submodel. 
    \item Search for lubricant menicus with the prescribed lubricant volume, using the lubricant meniscus submodel.
    \item Output the two-dimensional temperature solution for the lubricated ball-on-flat geometry, and heat rate at the lubricated contact, using the FEM heat conduction solver.
    \item Compute the total thermal conductance of the lubricated contact, i.e., \[\frac{\text{heat rate across lubricated contact}}{\text{temperature difference across lubricated contact}}\,.\]
\end{enumerate}

\subsection{Kogut-Etsion mechanical deformation model}\label{Kogut-Etsion mechanical deformation model}

\begin{figure}[H]
    \centering
    \includegraphics[width=0.7\linewidth]{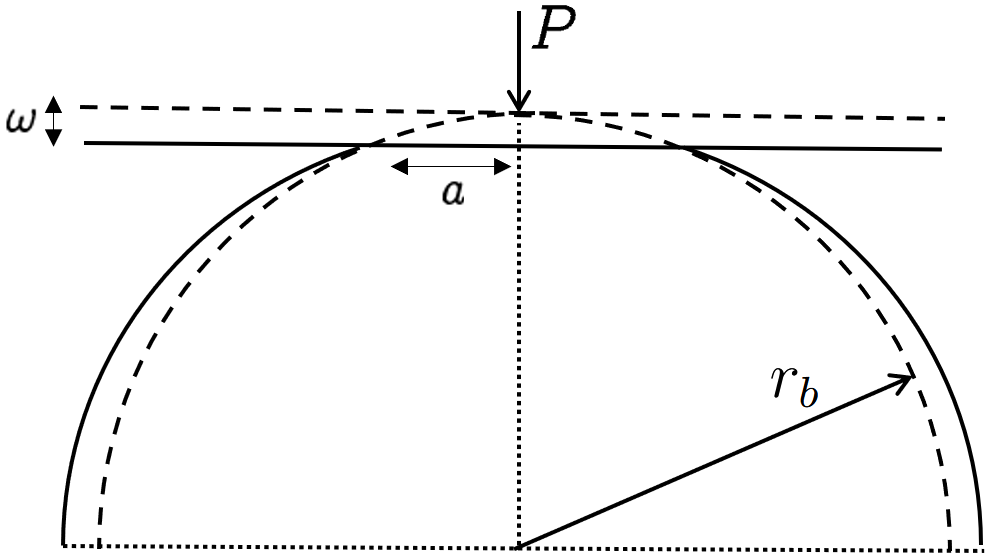}
    \caption{Deformable sphere on a rigid flat surface, modeled by Kogut and Etsion \cite{KogutEtsion2002}}
    \label{fig:DeformationModel}
\end{figure}

\indent The Kogut-Etsion model \cite{KogutEtsion2002} considers a deformable sphere pressed onto a rigid flat by an applied load. As shown in Fig. \ref{fig:DeformationModel}, the load \(P\) yields an interference \(\omega\) and contact radius \(a\). The load and contact radius relations follow the Hertz solution for \(0 < \omega \leq \omega_c\), where \(\omega_c\) is the critical interference \cite{Chang1988} given by
\begin{equation}\label{eq:CriticalInterference}
\omega_c = \ps*{\frac{\pi K_\text{ball} H_\text{ball}}{2 E_r}}^2 r_b
\end{equation}
where \(K_\text{ball} = 0.454 + 0.41 \nu\) is the ball's hardness coefficient dependent on its Poisson ratio \cite{Chang1988}, \( H_\text{ball} = 2.8 Y \) is the ball's hardness \cite{Tabor1951} dependent on its yield strength, and \(r_b\) is the ball radius. \(E_r = \ps*{\frac{1 - \nu_\text{ball}^2}{E_\text{ball}} + \frac{1 - \nu_\text{flat}^2}{E_\text{flat}}}^{-1}\) is the Hertz elastic modulus, where \(E_\text{ball}\) and \(E_\text{flat}\) are the respective materials' elastic moduli, and \(\nu_\text{ball}\) and \(\nu_\text{flat}\) are the respective materials' Poisson's ratios.\\
\indent For interferences that fit the range \(1 < \omega/\omega_c \leq 6\), the contact is elastic with a plastic region developing below the sphere surface, while in the range \(6 < \omega/\omega_c \leq 110\), the contact transitions from elastic-plastic to entirely plastic. From Kogut and Etsion's equations 2, 3, 7, and 8 \cite{KogutEtsion2002}, the interference ratio \(\omega/\omega_c\) can be expressed as a function of the applied load $P$, and in turn, the contact radius can be expressed as a function of the interference ratio as 
% \begin{align}
% & \frac{\omega}{\omega_c} = \ps*{\frac{P}{P_c}}^{2/3},\; \frac{a}{a_c} = \ps*{\frac{\omega}{\omega_c}}^{1/2} & \text{for }\; 0 < \frac{P}{P_c} \leq 1\\
% & \frac{\omega}{\omega_c} = 0.979\ps*{\frac{P}{P_c}}^{0.702},\; \frac{a}{a_c} = 0.964\ps*{\frac{\omega}{\omega_c}}^{0.880} & \text{for }\; 1 < \frac{P}{P_c} \leq 13.23\\
% & \frac{\omega}{\omega_c} = 0.766\ps*{\frac{P}{P_c}}^{0.792},\; \frac{a}{a_c} = 0.967\ps*{\frac{\omega}{\omega_c}}^{0.685} & \text{for }\; 13.23 < \frac{P}{P_c} \leq 530.16
% \end{align}
\begin{alignat}{2}
& \frac{\omega}{\omega_c} = \ps*{\frac{P}{P_c}}^\frac{2}{3},\; \frac{a}{a_c} = \ps*{\frac{\omega}{\omega_c}}^\frac{1}{2} && \quad \text{for }\; 0 < \frac{P}{P_c} \leq 1 \label{eq:InterferenceRatioBelow1}\\
& \frac{\omega}{\omega_c} = \ps*{\frac{1}{1.03} \frac{P}{P_c}}^\frac{1}{1.425},\; \frac{a}{a_c} = 0.93^\frac{1}{2}\ps*{\frac{\omega}{\omega_c}}^\frac{1.136}{2} && \quad \text{for }\; 1 < \frac{P}{P_c} \leq 13.23 \label{eq:InterferenceRatioAbove1}\\
& \frac{\omega}{\omega_c} = \ps*{\frac{1}{1.4} \frac{P}{P_c}}^\frac{1}{1.263},\; \frac{a}{a_c} = 0.94^\frac{1}{2}\ps*{\frac{\omega}{\omega_c}}^\frac{1.146}{2} && \quad \text{for }\; 13.23 < \frac{P}{P_c} \leq 530.16\label{eq:InterferenceRatioAbove13}
\end{alignat}
where \(P_c = \frac{4}{3}E_r r_b^{1/2} \omega_c^{3/2}\) and \(a_c = \ps*{r_b \omega_c }^{1/2}\). Once the contact radius $a$ due to the load \(P\) is calculated, it is input to the ball profile used by Yovanovich \cite{YovanovichKitscha1973}, for construction of the deformed ball geometry \(z_\text{ball}\ps*{r}\) used in the 2D azimuthally symmetric lubricant meniscus model. This deformed ball geometry neglects bulk deformation of the sphere, and is reasonable for small loads.
\begin{equation}\label{eq:YovanovichDeformedGeometry}
z_\text{ball}\ps*{r} = a\bs*{ \ps*{\frac{r_b}{a}}^2 - 1}^{1/2} - a\bs*{ \ps*{\frac{r_b}{a}}^2 - \ps*{\frac{r}{a}}^2}^{1/2}
\end{equation}

% Elastic from 0 to omega sub c of 1,
% Talk about critical interference omega sub c that marks transition from elastic to plastic regimes.\\
% Find load/critical load transition at omega star of 6, and upper boundary too.\\
% Include overall process to get to contact radius.

\subsection{2D Azimuthally symmetric lubricant meniscus model}
\begin{figure}[H]
    \centering
    \includegraphics[width=1.0\linewidth]{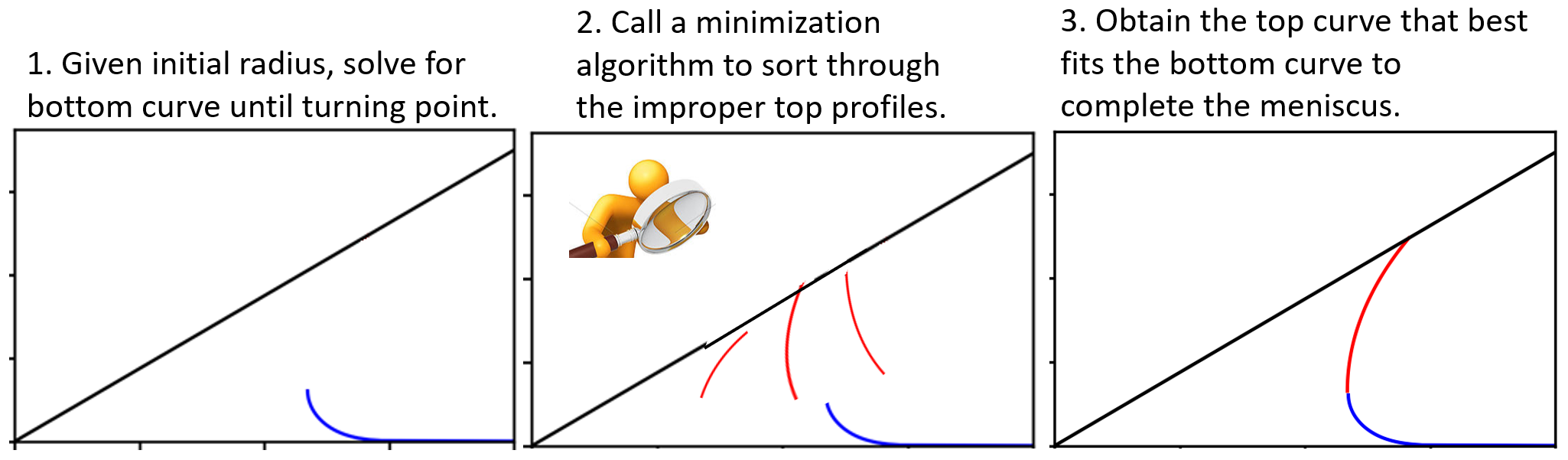}
    \caption{Flowchart of lubricant meniscus solver algorithm}
    \label{fig:MeniscusAlgorithm}
\end{figure}

\begin{figure}[H]
    \centering
    \includegraphics[width=0.6\linewidth]{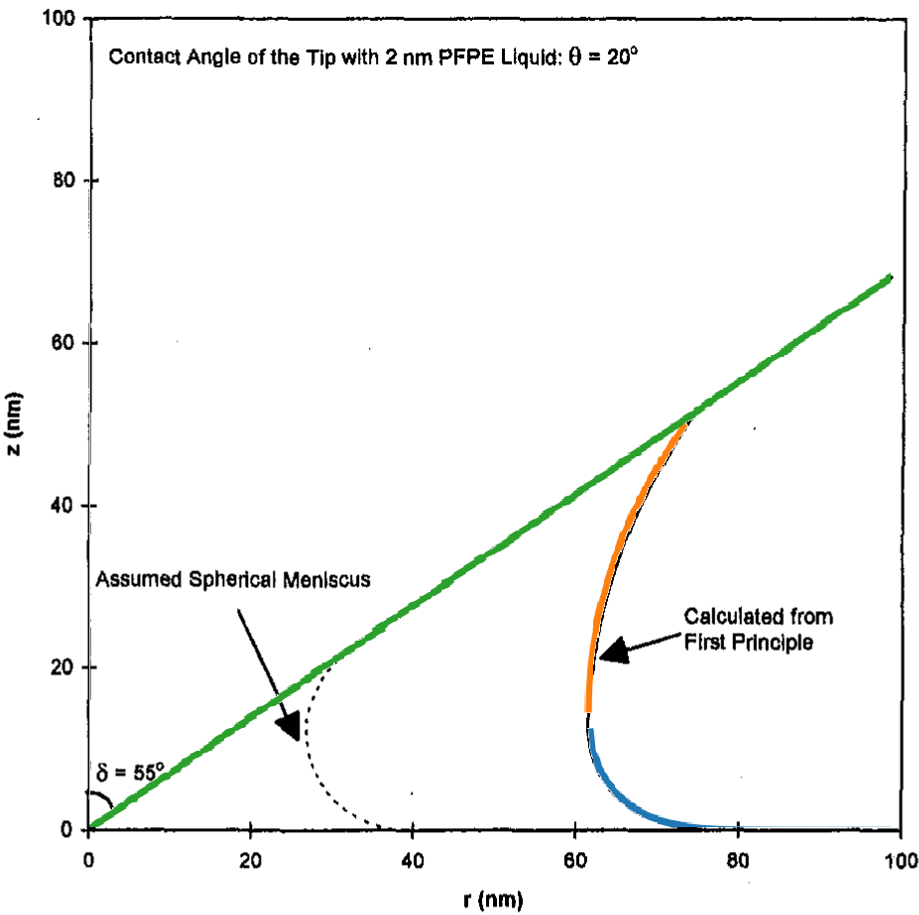}
    \caption{Our overlayed reproduction of Gao's result \cite{Gao1998} of a meniscus for a conical tip; reprinted with permission. Note that the orange and blue colors indicate ``top'' and ``bottom'' curves constructed separately.}
    \label{fig:MeniscusReproduction}
\end{figure}

For the liquid meniscus model, we employ the meniscus profile theory of Gao et al. \cite{Gao1998}. Their theory proposes the following differential equation to govern the pressure distribution within the meniscus.
\begin{equation}\label{eq:2DMeniscusDifferentialEquation}
\gamma_\text{LV} \ps*{\frac{\partial^2 z}{\partial x^2} + \frac{\partial^2 z}{\partial y^2}} + \rho g z = \frac{\alpha}{\ps*{z + t}^3} - \frac{\alpha}{z^3}
\end{equation}
\(x\) and \(y\) represent the in-plane, lateral dimensions, while \(z\) represents the vertical dimension; \(\gamma_\text{LV}\) is the surface tension of liquid on solid, \(\alpha\) is the surface interaction strength of the liquid, \(\rho\) is the density of the liquid, and \(g = 9.81\;\text{m}/\text{s}^2\) is the acceleration due to gravity. The differential equation Eq. \eqref{eq:2DMeniscusDifferentialEquation} balances the Laplace pressure due to local curvature (left side first term), the gravitational pressure (left side second term), and the disjoining pressure (right side terms). Upon applying azimuthal symmetry, the differential equation becomes
\begin{equation}\label{eq:2DMeniscusDEAzimuthalSymmetry}
    1 + \ps*{\hat{r}'}^2 - \hat{r} \hat{r}'' + \br*{\beta \hat{z} + \gamma \bs*{1 - \ps*{1 + \hat{z}}^{-3} }} \hat{r} \bs*{1 + \ps*{\hat{r}'}^2}^{3/2} = 0
\end{equation}
where \(t\) represents the liquid film thickness, \(\hat{z} = z/t\), \(\hat{r} = \ps*{x^2 + y^2}^{1/2}/t\), \(\hat{r}' = d\hat{r}/d\hat{z}\), \(\hat{r}'' = d^2\hat{r}/d\hat{z}^2\), \(\beta = \rho g t^2/\gamma_\text{LV}\), and \(\gamma = \alpha/\ps*{t^2 \gamma_\text{LV}}\).
To apply the above second order differential equation Eq. \eqref{eq:2DMeniscusDEAzimuthalSymmetry} in solving for the meniscus profile, we take the following solution approach (summarized in Fig. \ref{fig:MeniscusAlgorithm}), which we have used to replicate the meniscus profile shown by Gao et al. \cite{Gao1998} in Fig. \ref{fig:MeniscusReproduction}, as a means of illustrating the validity of our approach. 
\paragraph{\textbf{2D Lubricant Meniscus Solver}}
\begin{enumerate}
    \item Binary search is performed, to find the flat-lubricant intercept radius that yields a lubricant meniscus with the prescribed volume. Each iteration of this binary search requires the solution procedure to construct the meniscus profile guess, described in the next step. The volume of a lubricant meniscus for the ball-on-flat system is expressed as
    \begin{equation}\label{eq:2DMeniscusVolumeCalculation}
     V_\text{lubricant} = \pi \int_0^h \bs*{r^2_\text{meniscus}(z) - r^2_\text{ball}(z)} \, dz\;.
    \end{equation}
    \begin{figure}[H]
        \centering
        \includegraphics[width=0.25\linewidth]{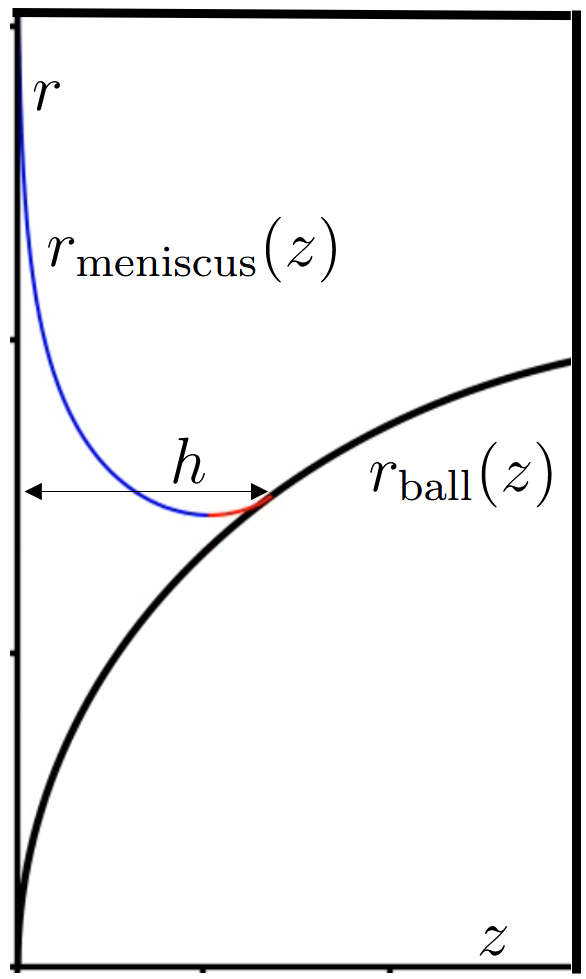}
        \caption{Visual aid for meniscus volume calculation by disc integration}
        \label{fig:2DMeniscusVolumeCalculation}
    \end{figure}
    \item For each flat-lubricant intercept guess \(r_\text{bottom}\), we solve for the meniscus profile \(z_b(r)\) that begins from the lubricant-flat intercept (referred to as the ``bottom'' curve, with subscript \textit{b}), and ends where the bottom curve \(z_b(r)\) reaches infinite slope, i.e., where \(\hat{r}' = d\hat{r}/d\hat{z} = 0\). This solution is accomplished by executing the explicit Runge-Kutta method of order 5(4) (\verb|scipy.integrate.solve_ivp| \cite{Virtanen2020}) on Eq. \eqref{eq:2DMeniscusDEAzimuthalSymmetry} decomposed into two first-order differential equations framed as an initial-value problem. A minimization approach (using the L-BFGS-B algorithm \cite{Virtanen2020}) is taken to discover the meniscus profile \(z_a(r)\) that begins at the ball-lubricant intercept (referred to as the ``top'' curve, with subscript \textit{a}) and ends exactly at the endpoint of the bottom curve \(z_b(r)\) while maintaining slope continuity. 
    \item Obtain the flat-lubricant intercept radius and corresponding meniscus profile that best contains the prescribed lubricant volume.
\end{enumerate}

\subsection{2D Heat transfer FEM solution}\label{2D Heat transfer FEM solution using FEniCS}
Once the lubricant meniscus geometry has been developed, we apply a steady state heat conduction solver to enable calculation of the total thermal conductance, through the following steps:
\begin{enumerate}
    \item Construct two-dimensional half-sphere, flat, and meniscus geometries symmetric about the vertical \(z\)-axis, and generate the corresponding mesh.
    \item Set material subdomains with specified thermal conductivities for the ball, flat, and lubricant meniscus. Set constant temperature boundaries at the top of the half-sphere and bottom of the flat, symmetry boundary condition at the \(z\)-axis, and insulated boundary at the vacuum interfaces.
    \item Solve the variational problem of steady state heat conduction in cylindrical coordinates using the FEniCS package \cite{Alnaes2015}. The weak form of the heat diffusion PDE can be expressed as
    \begin{equation}\label{eq:HeatDiffusionPDE}
     \int_\Omega -k \ps*{\frac{dT}{dr} \frac{dv}{dr} + \frac{dT}{dz} \frac{dv}{dz}} r\, dr = \int_\Omega 0 \cdot v \, dr,
    \end{equation}
    where \(v\) is the test function, \(k\) is the thermal conductivity of each material subdomain, and \(T\) is the trial function to obtain as a function of the radial and vertical position variables \((r, z)\). Fig. \ref{fig:BallonFlatSystem} illustrates the material subdomains, boundary conditions, and mesh density. Fig. \ref{fig:BallonFlatContours} shows examples of the temperature contour solution for both dry and lubricated cases. 
    \begin{figure}[H]
    \centering
    \begin{subfigure}{.45\textwidth}
        \centering
        \includegraphics[width=1\linewidth]{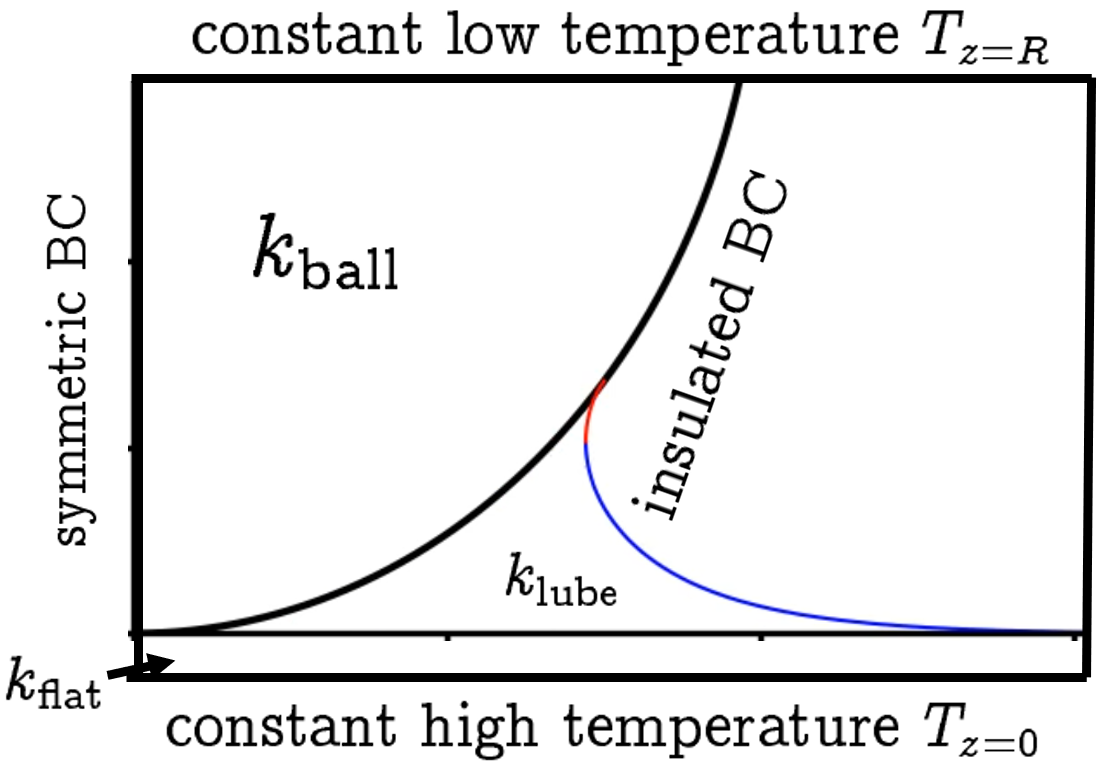}
        \caption{Ball-on-flat material subdomains\vspace{10px}}
        \label{fig:BallonFlatSubdomains}
    \end{subfigure}
    \hfill
    \begin{subfigure}{.54\textwidth}
        \centering
        \includegraphics[width=1\linewidth]{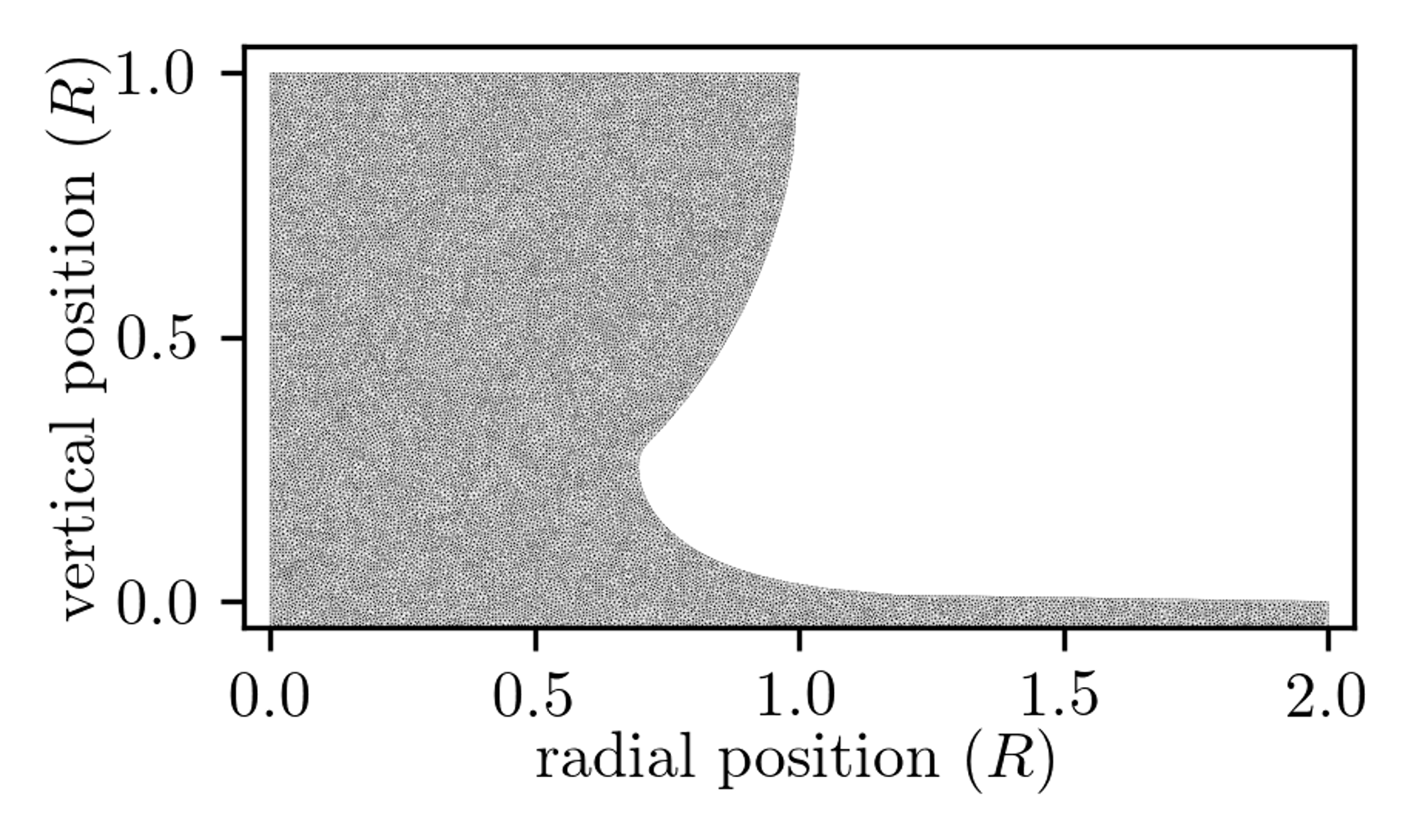}
        \caption{Ball-on-flat tetrahedral mesh}
        \label{fig:BallonFlatMesh}
    \end{subfigure}
    \caption{Ball-on-flat system described within Python code}
    \label{fig:BallonFlatSystem}
    \end{figure}
    
    \begin{figure}[H]
    \centering
    \begin{subfigure}{.41\textwidth}
        \centering
        \includegraphics[width=1\linewidth]{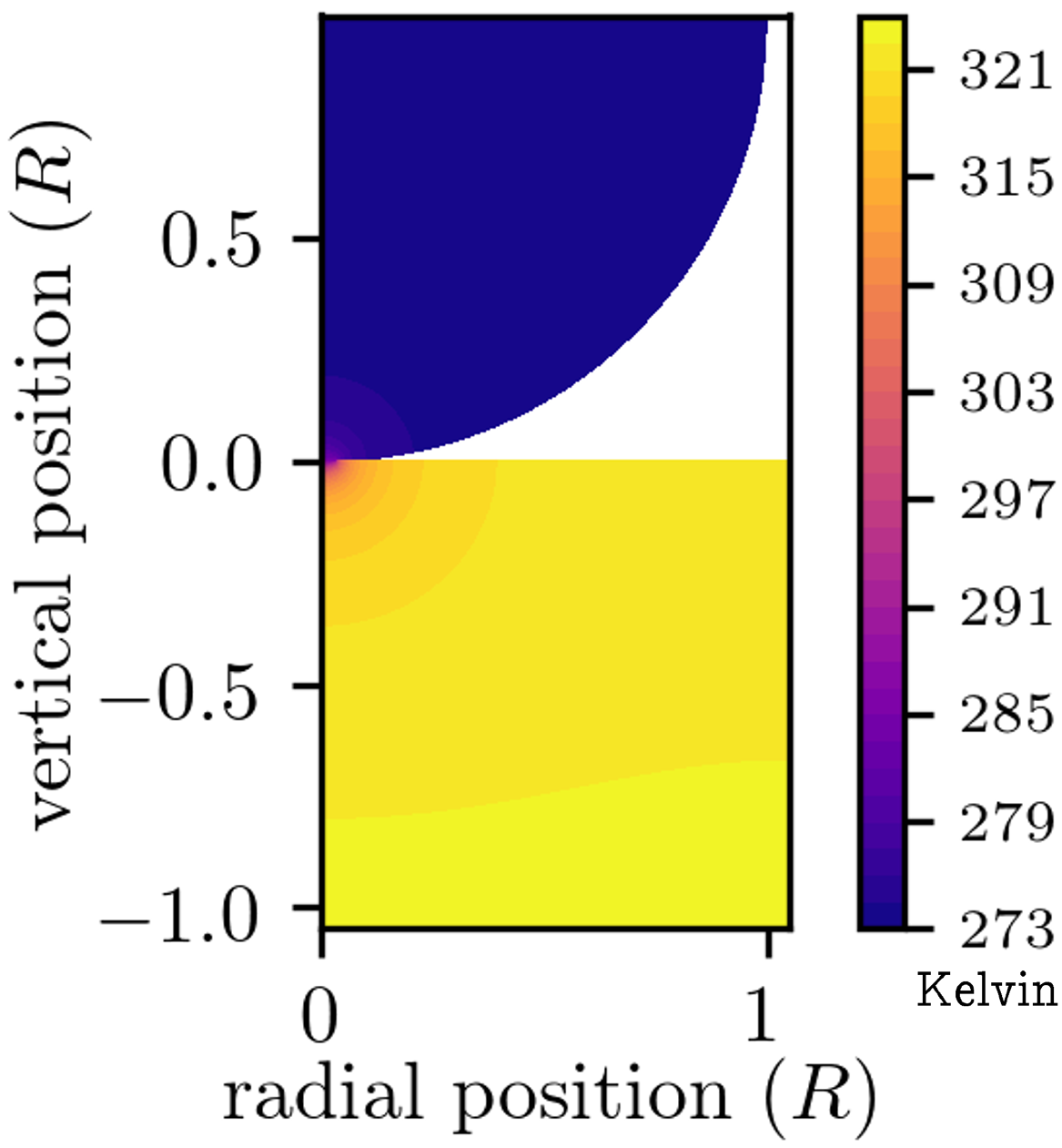}
        \caption{Dry ball-on-flat example temperatures}
        \label{fig:DryBallonFlatContours}
    \end{subfigure}
    \hfill
    \begin{subfigure}{.58\textwidth}
        \centering
        \includegraphics[width=1\linewidth]{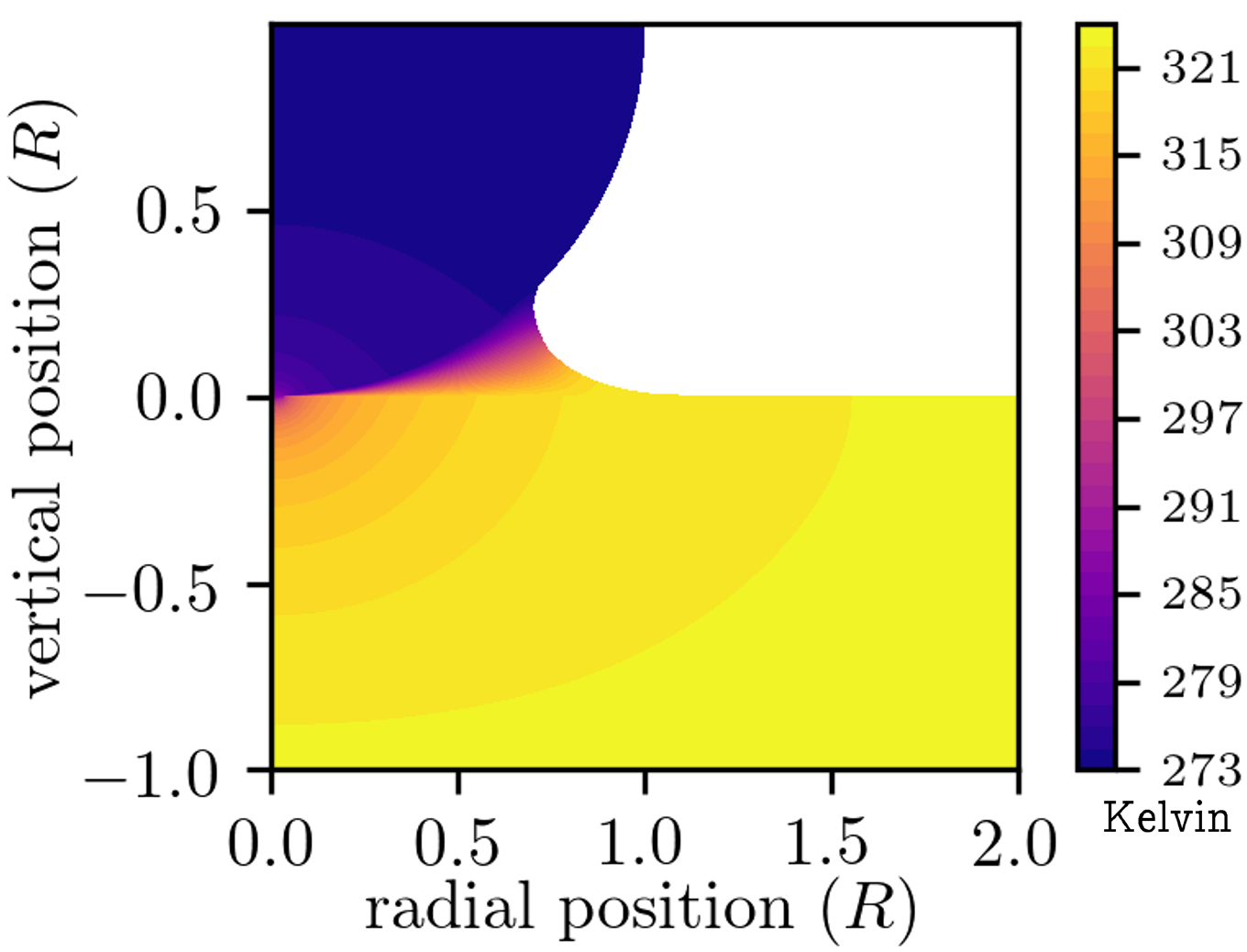}
        \caption{Lubricated ball-on-flat example temperatures\vspace{13px}}
        \label{fig:WetBallonFlatContours}
    \end{subfigure}
    \caption{Ball-on-flat example temperature contours, in which position is a fraction of ball radius \(R = r_b\)}
    \label{fig:BallonFlatContours}
    \end{figure}
    \item Upon obtaining the temperature solution, we calculate the total thermal conductance of the two-dimensional geometry assuming azimuthal symmetry by the following integral evaluated at the bottom of the race, where \(r\) is non-dimensionalized with respect to the ball radius \(r_b\): 
    \begin{equation}\label{eq:Gazimuthal2D}
         G_\text{azimuthal} = \frac{1}{T_\text{bottom of flat} - T_\text{top of ball}}\int_0^{r_\text{max}} -k_\text{flat} \frac{dT\ps*{r, z_\text{bottom of flat}}}{dz} \cdot 2\pi r r_b \, dr
     \end{equation}
     This calculation is separately performed for the same geometry but excluding the lubricant meniscus, to obtain the dry conductance of the two-dimensional, azimuthal system.
\end{enumerate}

\subsection{Correlation development using Yovanovich's numerical model}\label{Correlation development using Yovanovich's numerical model}
To serve as a comparison to the thermal resistance results of our multiphysics model of the ball-on-flat, we introduce Yovanovich's 1973 model of a lubricated ball-on-flat's thermal resistance \cite{YovanovichKitscha1973}. Yovanovich's model was based on two key assumptions: the ball-flat contact and the lubricant are considered to be uncoupled heat pathways, and the heat flows vertically in one-dimension through annular regions of the lubricant, which has a vertical wall for a meniscus.
\begin{figure}[H]
    \centering
    \includegraphics[width=0.5\linewidth]{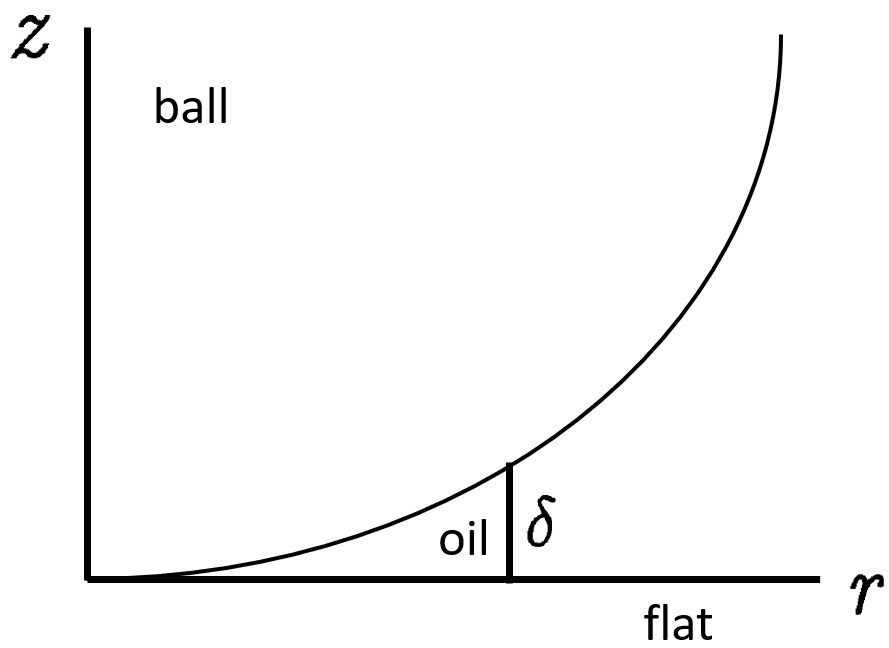}
    \caption{Yovanovich's model of a lubricated ball-on-flat, with vertical meniscus profile}
    \label{fig:YovanovichLubricantModel}
\end{figure}
Given these two assumptions, the differential thermal conductance of the lubricant is
\begin{equation}\label{eq:DifferentialThermalConductance}
     dG_l = \frac{k_l 2\pi r \, dr}{\delta(r)}
\end{equation}
where \(dG_l\) is the differential thermal conductance, \(k_l\) is the lubricant thermal conductivity, and \(\delta(r)\) is the lubricant thickness as a radial position \(r\). Also, the  ball-flat and ball-lubricant conductances sum in parallel as
\begin{equation}\label{eq:ConductanceSum}
     G_\text{total} = G_\text{dry} + G_\text{lub}\,,
\end{equation}
where \(G_\text{dry}\) is the thermal conductance of the dry ball-on-flat, and \(G_\text{lub}\) is the thermal conductance of the lubricated ball-on-flat system. From Eq. \eqref{eq:DifferentialThermalConductance} it follows that the total lubricant conductance is computed by the following integral.
\begin{align}
    G_\text{lub} &  = \int_{r_\text{min}}^{r_\text{wet}} \frac{k_l 2\pi r \, dr}{\sqrt{r_b^2 - a_\text{H}^2} - \sqrt{r_b^2 - r^2}}\label{eq:ConductanceIntegral}\\
    & = 2\pi k_l \bs*{\sqrt{r_b^2 - a_\text{H}^2} \ln\abs*{\sqrt{r_b^2 - a_\text{H}^2} - \sqrt{r_b^2 - r^2}} + \sqrt{r_b^2 - r^2}}_{r_\text{min}}^{r_\text{wet}}, \label{eq:ConductanceIntegralEvaluated}
\end{align}
where \(a_\text{H}\) is the contact radius that results for loading in a Hertzian contact:
\begin{equation}\label{eq:HertzianContactRadius}
     a_\text{H} = \ps*{\frac{3 r_b P}{4 E_r}}^{1/3}\,.
\end{equation}
The wetted radius \(r_\text{wet}\) is obtained by numerically inverting
\begin{equation}\label{eq:VolumeForAbitraryWettedRadius}
     V_l = \pi\bs*{r_l^2 \sqrt{r_b^2 - a_\text{H}^2} + \frac{2}{3}\ps*{r_b^2 - r_l^2}^{3/2} - a_\text{H}^2\sqrt{r_b^2 - a_\text{H}^2} - \frac{2}{3}\ps*{r_b^2 - a_\text{H}^2}^{3/2}}\;\text{at }r_l = r_\text{wet}\,.
\end{equation}
% From Yov approximation of lubricant geometry! Need to explain that.
The minimum lubricant radius \(r_\text{min}\) imposes the condition of Knudsen number Kn \(<\) 0.01 for continuum flow, i.e.,
\begin{equation}\label{eq:MinRadiusForMaxKnudsen}
     r_\text{min} = \sqrt{r_b^2 - \ps*{\sqrt{r_b^2 - a_\text{H}^2} - \frac{\text{lubricant intermolecular distance}}{\text{Kn}_\text{max}}}^2}, \text{ where } \text{Kn}_\text{max} = 0.01.
\end{equation}
As Yovanovich's model of thermal conductance of the lubricated ball-on-flat system does not have a closed, analytical form due to requiring the aforementioned calculation steps, we investigated the effect of varying the input variables on the output thermal conductance, to develop a correlation equation that encapsulates the relationship between the inputs and outputs. A significant motivation for this correlation development is that we aim to construct a correlation equation for our multiphysics based ball-on-flat model as well and then compare the correlations. These correlations may be useful to scientific and engineering communities that wish to better understand ball bearing thermal conductance, both from a pure physics standpoint and for experimental or hardware design purposes.

To begin explaining our approach to correlation development, we reiterate that the core issue in Yovanovich's numerical model that prevents a closed analytical solution to total thermal conductance is that Eq. \eqref{eq:ConductanceIntegralEvaluated} for \(G_\text{lub}\) depends on the wetted radius \(r_\text{wet}\), which must be calculated by inverting Eq. \eqref{eq:VolumeForAbitraryWettedRadius} as the lubricant volume \(V_l\) depends on \(r_\text{wet}\). Let us rewrite the integral result in Eq. \eqref{eq:ConductanceIntegralEvaluated} in its indefinite form, without the constant prefactor \(2\pi k_l\) as follows.
\begin{equation}\label{eq:YovanovichIntegralResultIndefinite}
     I_\text{Yovanovich}\ps*{r;\,r_b, a_\text{H}} = \sqrt{r_b^2 - a_\text{H}^2} \ln\abs*{\sqrt{r_b^2 - a_\text{H}^2} - \sqrt{r_b^2 - r^2}} + \sqrt{r_b^2 - r^2}
\end{equation}
Upon inspecting graphically how \(I_\text{Yovanovich}\) of Eq. \eqref{eq:YovanovichIntegralResultIndefinite} indirectly depends on the lubricant volume \(V_l\), we found that the following correlation \(I_\text{fit, Yov.}\) is able to fit the numerical results of \(I_\text{Yovanovich}/r_b\).
\begin{equation}\label{eq:YovanovichCorrelationFit}
    I_\text{fit, Yov.}\ps*{V_l;\,r_b, a_\text{H}} = \ps*{c_0 \ln{\frac{a_\text{H}}{r_b}} + c_1} \ln\ps*{\frac{V_l}{\frac{4}{3} \pi r_b^3}} + c_{2} \ln{\frac{a_\text{H}}{r_b}} + c_{3}
\end{equation}
The constants \(c_0 \dots c_{3}\) are discovered by first performing linear regression to discover the ``slope'' that multiplies onto \(\ln\ps*{\frac{V_l}{\frac{4}{3} \pi r_b^3}}\) in Eq. \eqref{eq:YovanovichCorrelationFit} and ``intercept'' which is the sum of the second and third terms in Eq. \eqref{eq:YovanovichCorrelationFit}, for combinations of ball radius \(r_b\), load \(P\), and Hertz elastic modulus \(E\). Next, these slopes and intercepts are fit to the slope function \(\ps*{c_0 \ln{\frac{a_\text{H}}{r_b}} + c_1}\) and intercept function \(\ps*{c_{2} \ln{\frac{a_\text{H}}{r_b}} + c_{3}}\), respectively, by applying nonlinear least-squares with the Trust Region Reflective algorithm\\ (\verb|scipy.optimize.least_squares(method=`trf')|\cite{Virtanen2020}). The slope and intercept terms capture the dependence of \(I_\text{Yovanovich}/r_b\) on the ball radius \(r_b\) and contact radius \(a_\text{H}\), while the \(\ln\ps*{\frac{V_l}{\frac{4}{3} \pi r_b^3}}\) term captures the dependence on lubricant volume \(V_l\). Note that while \(I_\text{fit, Yov.}\) is made dimensionless, the expression \(r_b I_\text{fit, Yov.}\) maintains units of length, as the dependence on \(r_b\) manifests from the integral that gives rise to the total lubricant conductance (Eq. \eqref{eq:ConductanceIntegralEvaluated}). Now, given Yovanovich's equation for the dry thermal conductance of a circular contact \cite{Yovanovich1967},
\begin{equation}\label{eq:DryThermalConductance}
     G_\text{dry} = 4a_\text{H}\ps*{\frac{1}{k_f} + \frac{1}{k_b}}^{-1}
\end{equation}
where \(k_f\) and \(k_b\) are the flat and ball thermal conductivities respectively, the ratio of total thermal conductance to dry conductance can then be expressed as
\begin{equation}\label{eq:ConductanceRatioYovanovich}
     \frac{G_\text{total, Yov.}}{G_\text{dry, Yov.}} = \frac{\pi k_l r_b}{2 a_\text{H} k_r} \bs*{I_\text{fit, Yov.}\ps*{V_\text{wet};\,r_b, a_\text{H}} - I_\text{fit, Yov.} \ps*{V_\text{min};\,r_b, a_\text{H}}} + 1,
\end{equation}
where \(k_r = \ps*{1/k_f + 1/k_b}^{-1}\) is the reduced thermal conductivity of the ball-flat interface, and 
\begin{align}
    V_\text{wet} & = \pi\bs*{r_\text{wet}^2 \sqrt{r_b^2 - a_\text{H}^2} + \frac{2}{3}\ps*{r_b^2 - r_\text{wet}^2}^{3/2} - a_\text{H}^2\sqrt{r_b^2 - a_\text{H}^2} - \frac{2}{3}\ps*{r_b^2 - a_\text{H}^2}^{3/2}},\label{eq:VolumeMaximum}\\
    V_\text{min} & = \pi\bs*{r_\text{min}^2 \sqrt{r_b^2 - a_\text{H}^2} + \frac{2}{3}\ps*{r_b^2 - r_\text{min}^2}^{3/2} - a_\text{H}^2\sqrt{r_b^2 - a_\text{H}^2} - \frac{2}{3}\ps*{r_b^2 - a_\text{H}^2}^{3/2}},\label{eq:VolumeMinimum}
\end{align}
which follow from Eq. \eqref{eq:VolumeForAbitraryWettedRadius}.
% We notice how the conductance ratio features k_l/k_r as a coefficient, which will carryover to the 2D/3D correlations as we will adopt similar correlation structure.

% \subsection{Ball-in-races model}
% \subsubsection{3D lubricant meniscus model}
% \subsubsection{Ball-in-races finite element model in Abaqus}

\section{Methods for ball-on-races model}
\label{Methods for ball-on-races model}
In addition to developing correlations from our multiphysics ball-on-flat model and Yovanovich's lubricated ball-on-flat model, we would like to establish a comparison between a lubricated ball-on-flat's thermal resistance and a lubricated angular contact bearing's thermal resistance. The motivation for this is that the ball-on-flat geometry can be quickly evaluated via numerical analysis and experimental tests, as the geometry is less complex to set up and evaluate (both computationally and experimentally) compared to the angular contact ball bearing geometry. By developing a numerical understanding of the discrepancy between the ball-on-flat thermal conductance and a ball-on-races conductance, we can make educated predictions of what a ball-on-flat experimental test result would imply for the conductance of an actual static bearing.

\subsection{3D Mechanical deformation model}\label{3D mechanical deformation model}

To model the three-dimensional deformation, we again make use of the Kogut-Etsion model as in Section \ref{Kogut-Etsion mechanical deformation model}. However, as the three-dimensional ball-on-race system is not azimuthally symmetric with not only a single contact radius like in the two-dimensional ball-on-flat system, we adjust the implementation to incorporate the interference \(\omega\) directly. Specifically, the ball is displaced by the interference \(\omega\) into the race to create the intersection between the ball and race, where the region of ball displaced into the race is truncated away. This technique of modeling the ball deformation yields a similar geometry to that of Yovanovich's model used in Section \ref{Kogut-Etsion mechanical deformation model}, which is reasonable for loads that would not cause significant bulk deformation of the sphere. The steps to generate the ball-race intersection radii follows.
\paragraph{\textbf{3D Contact Ellipse Solver}}
\begin{enumerate}[leftmargin=0pt]
    \item Calculate the interference \(\omega\) to displace the ball into the race, using the Kogut-Etsion model in Eqs. \eqref{eq:InterferenceRatioBelow1}-\eqref{eq:InterferenceRatioAbove13}.
    \item Construct the ball geometry \(z_b\ps*{r, \varphi}\) and race geometry \(z_r\ps*{r, \varphi}\). The outer race geometry requires no transformation as we assume that it is not mechanically deformed, and it is expressed as a section of the following torus:
    \begin{numcases}{z_r\ps*{r, \varphi}=}
    -\sqrt{ -\bs*{r \cos\ps*{\varphi}}^2+ \br*{\sqrt{r_r^2 - \bs*{r \sin\ps*{\varphi} - r_r  \sin\ps*{\varphi_b}}^2} + r_\text{bearing}}^2 } + r_\text{bearing} + r_r \nonumber \\
    \hfill \text{for } r\sin\ps*{\varphi} < r_r\sin\ps*{\varphi_b} \\
    -\sqrt{\ps*{r_\text{bearing} + r_r}^2 - \bs*{r \cos\ps*{\varphi}}^2} + r_\text{bearing} + r_r \nonumber \\
    \hfill \text{for } r\sin\ps*{\varphi} \geq r_r\sin\ps*{\varphi_b}
    \end{numcases}
    where \(r_\text{bearing}\) is the distance from the bearing ring center to the ball center and \(\varphi_b\) is the bearing angle. The ball geometry is made by first creating a sphere surface that perfectly meets the junction of the curved and flat race surfaces of \(z_r\ps*{r, \varphi}\), before rotating it by the bearing angle \(\varphi_b\) and translating it into the race by the  Kogut-Etsion interference \(\omega\):
    \begin{align}\label{eq:3DBallGeo}
        x_{b} & = u \cos\ps*{\varphi}, \text{ for } 0 \leq u \leq r_b \text{ and } -\pi/2 \leq \varphi \leq \pi/2\\
        y_{b,\,\text{original}} & = u \sin\ps*{\varphi}\\
        z_{b,\,\text{original}} & = r_b - \sqrt{r_b^2 - u^2}\\
        y_{b,\,\text{transform}} & = y_{b,\,\text{original}} \cos\ps*{\varphi_b} + z_{b,\,\text{original}} \sin\ps*{\varphi_b}
        - \omega \sin \ps*{\varphi_b}\\
        z_{b,\,\text{transform}} & = -y_{b,\,\text{original}} \sin\ps*{\varphi_b} + z_{b,\,\text{original}} \cos\ps*{\varphi_b} + r_r \bs*{1 - \cos\ps*{\varphi_b} } - \omega \cos\ps*{\varphi_b}\\
        r & = \sqrt{x_{b}^2 + y_{b,\,\text{transform}}^2}\\
        z_b\ps*{r, \varphi} & = \text{scipy.interpolate.PchipInterpolator}\ps*{r,\; z_{b,\,\text{transform}}}
    \end{align}
    where \(z_b\ps*{r, \varphi}\) is output as a Piecewise Cubic Hermite Interpolating Polynomial (PCHIP) \cite{Virtanen2020}, a callable function that maps \(r\) to \(z_{b,\,\text{transform}}\).
    \item Solving for \(r\ps*{\varphi}\) where \(z_r\ps*{r, \varphi} = z_b\ps*{r, \varphi}\) yields the contact radii \(r_\text{contact}\ps*{\varphi}\) due to the load-driven displacement \(\omega\). Taking \(b_{\text{con},\,+} = r_\text{contact}\ps*{\pi/2}\) and \(b_{\text{con},\,-} = r_\text{contact}\ps*{-\pi/2}\) yields the contact semi-minor axes' lengths in the positive and negative directions respectively, while \(a_\text{con} = r_\text{contact}\ps*{0}\) is the semi-major axis length. 
\end{enumerate}

\subsection{3D Lubricant meniscus model}

\begin{figure}[H]
\centering
\begin{subfigure}{.49\textwidth}
    \centering
    \includegraphics[width=1\linewidth]{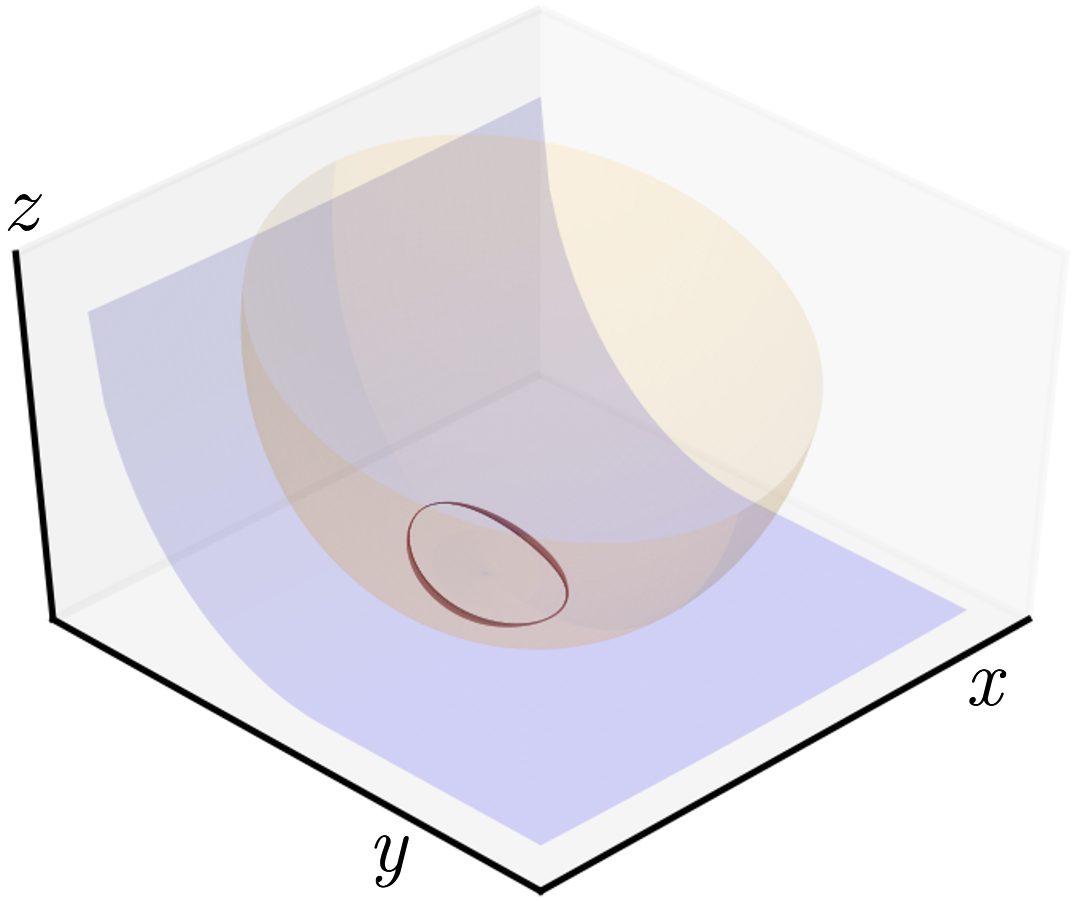}
    \caption{\SI{0.5}{\micro\liter} lubricant volume}
    \label{fig:3DSmallMeniscus}
\end{subfigure}
\hfill
\begin{subfigure}{.49\textwidth}
    \centering
    \includegraphics[width=1\linewidth]{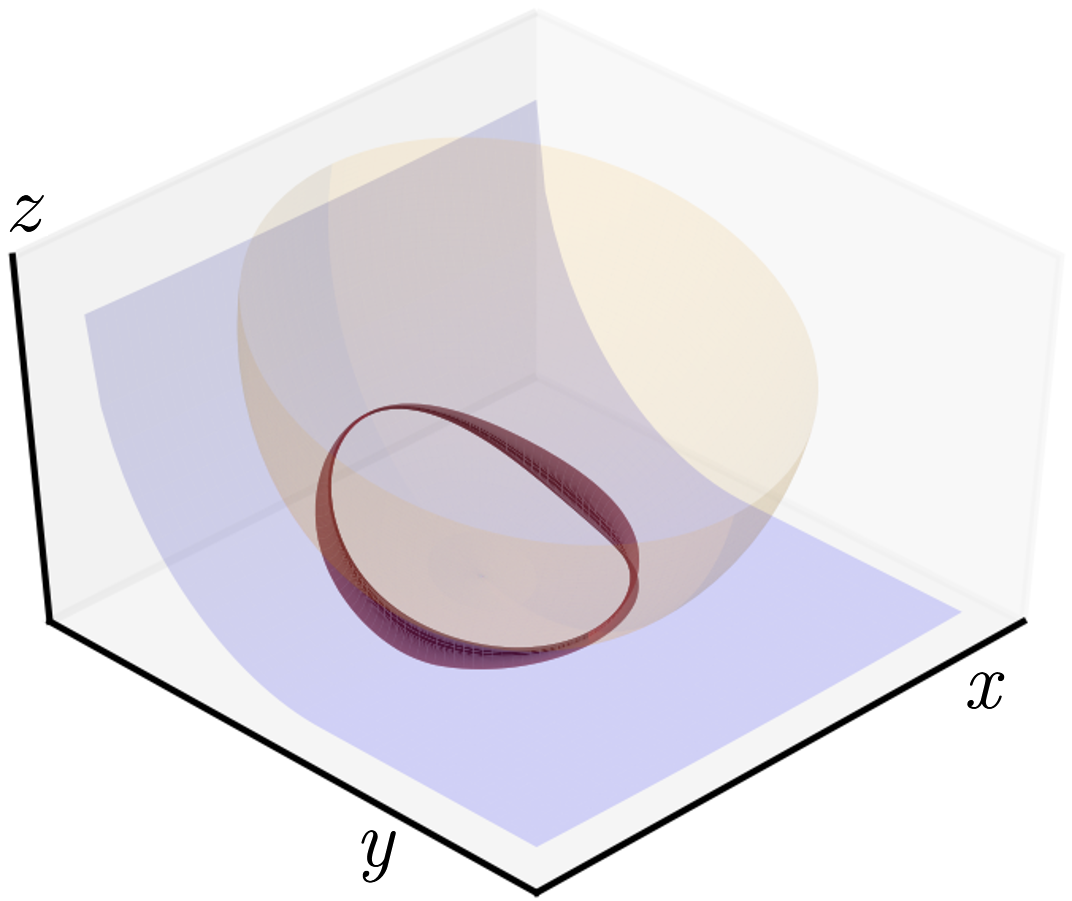}
    \caption{\SI{10}{\micro\liter} lubricant volume}
    \label{fig:3DLargeMeniscus}
\end{subfigure}
\caption{Example lubricant meniscus solutions for a \SI{1}{\centi\meter} radius ball}
\label{fig:3DMeniscii}
\end{figure}

Similar to the two-dimensional lubricant meniscus model, we utilize the meniscus profile theory of Gao et al. \cite{Gao1998}, and frame the process of solving for the three-dimensional lubricant meniscus as a minimization problem. Specifically, we must minimize two residuals simultaneously, i.e., a pressure balance residual and a volume residual both in cylindrical coordinates, where the \(x\) and \(z\) directions are depicted in Fig. \ref{fig:3DMeniscii} and the radial position \(r\) begins at the center of the ball-race contact:
\begin{equation}\label{eq:3DMeniscusMinimization}
    \min \quad \abs*{\frac{d^2 x}{dr^2} + \frac{1}{r}\frac{dz}{dr} + \frac{1}{r^2} \frac{d^2 z}{d\varphi^2} - \frac{1}{\gamma_\text{LV}}\bs*{ \frac{\alpha}{\ps*{z + t}^3} - \frac{\alpha}{z^3} - \rho g z}} \times \abs*{V_\text{input} - \iiiint r\, dr\, dz\, d\varphi}\,.
\end{equation}
\noindent Note that the pressure balance residual is based directly on Eq. \eqref{eq:2DMeniscusDifferentialEquation}, in which we take the \textit{x}-axis to be orthogonal to the \textit{yz}-plane that divides the three-dimensional meniscus into two symmetric halves. Two key assumptions are made about the lubricant meniscus solution. First, we assert that \(r\ps*{\varphi}\) is elliptical, centered at the ball-race contact point, with two possibly different semi-minor axis lengths (in which the semi-minor axes are in the \textit{y} direction). Also, to simplify the pressure balance residual calculation, we take the profile \(z\bs*{r\ps*{\varphi}}\) of each meniscus slice at azimuthal angle \(\varphi\) to be a circular arc. As a side note, for the illustrations that follow, we treat the ball as sitting on an outer race in which gravity points downward, but our methodology also supports gravity pointing the opposite direction which allows modeling the lubricant meniscus of a ball on an inner race as well. The algorithm to solve for the three-dimensional meniscus based on the pressure balance and volume residuals is as follows.
\paragraph{\textbf{3D Lubricant Meniscus Solver}}
\begin{enumerate}
    \item Define a function that takes input variables of the semi-major axis length \(a_\text{lub}\) (along the \textit{x}-axis) and semi-minor axes' lengths \(b_{\text{lub},\,+}\) and \(b_{\text{lub},\,-}\) (in the positive and negative \textit{y}-axis directions, respectively) of the ellipse that radially bounds the meniscus, and outputs the radial position \(r_\text{ellipse}\ps*{\varphi}\) of each meniscus slice at arbitrary azimuthal angle \(\varphi\) where $-\pi/2 \leq \varphi \leq \pi/2$.
    \item Define a function that calculates the two-dimensional, circular meniscus profile \(z_\text{meniscus, 2D}\ps*{r}\) for a particular azimuthal angle \(\varphi\), where the minimum radius of the profile is located at \(r_\text{ellipse}\ps*{\varphi}\) defined in Step 1.
    \item Define an objective function in which the input variables are the semi-major axis length \(a\) and semi-minor axes' lengths \(b_{\text{lub},\,+}\) and \(b_{\text{lub},\,-}\) of the meniscus, and the objective \textit{f} to minimize is the product of the pressure balance and volume residuals which are functions of the three-dimensional \(z_\text{meniscus, 3D}\bs*{r\ps*{\varphi}}\).
    \item Apply Nelder-Mead minimization [\verb|scipy.optimize.minimize(method=`Nelder-Mead')| \cite{Virtanen2020}] to the objective function, to obtain a three-dimensional lubricant meniscus that both fulfills the desired volume as well as meets the pressure balance constraint as much as possible. 
\end{enumerate}
Fig. \ref{fig:3DMeniscii} illustrates two representative meniscii for lubricant volumes of \SI{0.5}{\micro\liter} and \SI{10}{\micro\liter}, for a \SI{1}{\centi\meter} radius ball.

\subsection{3D Heat transfer FEM solution based on 2D meniscus cross-sections}
Once we have generated a three-dimensional lubricant geometry, we can evaluate the effect of lubricant on the thermal conductance of a ball-on-races system by repurposing elements of the approach used for the 2D azimuthal geometry. In essence, we can apply the steady state heat conduction solver in cylindrical coordinates (using FEniCS) to two-dimensional cross-sections of the lubricant meniscus at many azimuthal angles \(\varphi\), and perform a weighted sum over them where the weights account for the race and ball surface areas occupied by a meniscus slice at angle \(\varphi\) in both the actual ball-on-race geometry and the imagined azimuthally symmetric ball-on-race geometry for a certain meniscus slice. While this quasi-3D approach yields lower computational cost compared to 3D heat transfer solution in a full-fledged FEM engineering software package, it has the physics limitation of not accurately capturing heat transfer in the azimuthal angle direction \(\varphi\) which cannot be avoided even with higher resolution of discretization in angle \(\varphi\). We detail the solution procedure below.
\paragraph{\textbf{3D Heat Transfer Solver}}
\begin{enumerate}
    \item For the meniscus cross-section at azimuthal angle \(\varphi\) for \(-\pi/2\leq\phi\leq\pi/2\), we construct the corresponding two-dimensional half-sphere, race, and circular meniscus profile. As explained in Section \ref{2D Heat transfer FEM solution using FEniCS} on the 2D heat transfer FEM solution, we calculate the total thermal conductance of the two-dimensional geometry assuming azimuthal symmetry by the following integral evaluated at the bottom of the race: 
    \begin{equation}\label{eq:Gazimuthal3D}
         G_\text{azimuthal} = \frac{1}{T_\text{bottom of race} - T_\text{top of ball}}\int_0^{r_\text{max}} -k_\text{race} \frac{dT\ps*{r, z_\text{bottom of race}}}{dz} \cdot 2\pi r r_b \, dr
     \end{equation}
     Again, this calculation is separately performed for the same geometry but excluding the lubricant meniscus, to obtain the dry conductance of the azimuthal system.
     \item Each conductance at azimuthal angle \(\varphi\) is multiplied by the following weight:
     \begin{align} \label{eq:MeniscusWeight}
         \begin{split}
            \text{weight}\ps*{\varphi} & =  2\Delta \varphi \cdot \text{mean}\Biggl\{\int_0^{r_\text{max}} r r_b \sqrt{1 + \bs*{\frac{\partial z_\text{ball}\ps*{r, \varphi}}{\partial r}}^2 + \frac{1}{r}\bs*{1 + \frac{\partial z_\text{ball}\ps*{r, \varphi}}{\partial \varphi}}^2 }\,dr,\\ & \int_0^{r_\text{max}} r r_b \sqrt{1 + \bs*{\frac{\partial z_\text{race}\ps*{r, \varphi}}{\partial r}}^2 + \frac{1}{r}\bs*{\frac{\partial z_\text{race}\ps*{r, \varphi}}{\partial \varphi}}^2 }\,dr \Biggr\} \\& \Bigg/ \br*{ 2\pi \cdot \text{mean}\br*{ \int_0^{r_\text{max}} r r_b \sqrt{1 + \bs*{z_\text{ball}'\ps*{r} }^2 }\,dr,\; \int_0^{r_\text{max}} r r_b \sqrt{1 + \bs*{z_\text{race}'\ps*{r} }^2 }\,dr }}\;,
        \end{split}
    \end{align}
    which is the average of the surface areas of the actual meniscus slice (covering angle \(\Delta \varphi\)) on the ball and race, divided by the average of the surface areas of the meniscus (covering angle \(2\pi\)) on the ball and flat in the azimuthally symmetric system.
    \item Once we have the weighted conductance at each azimuthal angle \(\varphi\), they are summed together to obtain the total conductance of the three-dimensional meniscus:
    \begin{equation} \label{eq:Gactual}
        G_\text{total, 3D} = \sum_{\varphi \in \ps*{-\frac{\pi}{2},\,\frac{\pi}{2}}} \text{weight}\ps*{\varphi} \; G_\text{azimuthal}\ps*{\varphi}
    \end{equation}
\end{enumerate}

% \[\text{weight}\ps*{\varphi} = \frac{\text{ball-race surface area of meniscus slice (covering angle } \Delta \varphi)}{\text{ball-flat surface area of meniscus (covering angle } 2 \pi)}\]

\subsection{Correlation development for 2D and 3D multiphysics models} \label{Correlation development for 2D and 3D multiphysics models}

Similarly to how we developed a thermal conductance correlation for Yovanovich's model in Section \ref{Correlation development using Yovanovich's numerical model}, we can also develop correlations for the two-dimensional and three-dimensional multiphysics models, and compare them to that for Yovanovich's model. To enable effective comparison and mirror the motivation underlying Yovanovich's model, we start with the structure of Eqs. \eqref{eq:YovanovichCorrelationFit} and \eqref{eq:ConductanceRatioYovanovich}, but modify the calculation of the contact radius \(a\) in Eq. \eqref{eq:ConductanceRatioYovanovich} to be model-specific and remove the \(I_\text{fit}\ps*{V_\text{min}}\) term in Eq. \eqref{eq:YovanovichCorrelationFit} which is not motivated in our custom 2D and 3D multiphysics models.

For the two-dimensional multiphysics model, the thermal conductance ratio from Eq. \eqref{eq:ConductanceRatioYovanovich} is modified to
\begin{equation}\label{eq:ConductanceRatio2D}
     \frac{G_\text{total, 2D multi.}}{G_\text{dry, 2D}} = \frac{\pi k_l r_b}{2 a_\text{2D} k_r} I_\text{fit, 2D}\ps*{V_l; r_b, a_\text{2D}} + 1,
\end{equation}
where \(a_\text{2D}\) is obtained from the Kogut-Etsion model in Eqs. \eqref{eq:InterferenceRatioBelow1}-\eqref{eq:InterferenceRatioAbove13}, and \(I_\text{fit, 2D}\) is based on Eq. \eqref{eq:YovanovichCorrelationFit}, replacing \(a_\text{H}\) with \(a\) specific to either the 2D or 3D model as follows:
\begin{equation}\label{eq:2D3DCorrelationFit}
    I_\text{fit, 2D or 3D, 4}\ps*{V_l;\,r_b, a} = \ps*{c_0 \ln{\frac{a}{r_b}} + c_1} \ln\ps*{\frac{V_l}{\frac{4}{3} \pi r_b^3}} + c_{2} \ln{\frac{a}{r_b}} + c_{3}\,.
\end{equation}
Additionally, for the three-dimensional multiphysics model, we will compare the accuracy of the above correlation structure with 4 fitting constants to that of the following correlation structure with 6 fitting constants:
\begin{equation}\label{eq:3DCorrelationFit2}
    I_\text{fit, 3D, 6}\ps*{V_l;\,r_b, a} = \ps*{c_0 \ln{\frac{a}{r_b}} + c_1} \ln^2\ps*{\frac{V_l}{\frac{4}{3} \pi r_b^3}} + \ps*{c_{2} \ln{\frac{a}{r_b}} + c_{3}} \ln\ps*{\frac{V_l}{\frac{4}{3} \pi r_b^3}} + c_{4} \ln{\frac{a}{r_b}} + c_{5}\,.
\end{equation}
Then, for the three-dimensional multiphysics model, the thermal condunctance ratio is analogously modeled as
\begin{equation}\label{eq:ConductanceRatio3D}
     \frac{G_\text{total, 3D multi.}}{G_\text{dry, 3D}} = \frac{\pi k_l r_b}{2 a_\text{3D} k_r} I_\text{fit, 3D}\ps*{V_l; r_b, a_\text{3D}} + 1,
\end{equation}
where \(a_\text{3D}\) is chosen to be the radius of the circle with the same area as the contact ellipse, which is computed by taking the geometric mean of the semi-major and semi-minor axes below. As there are two semi-minor axes, the mean of the semi-minor axes is taken first before computing the geometric mean. While other expressions for \(a_\text{3D}\) that encapsulate the varying contact radii of the contact ellipse could also be chosen in constructing correlations, we chose this one as it is an analogue of the contact area, whereas, for example, a simple mean would not be.
\begin{equation}\label{eq:ContactRadiusEllipse}
     a_\text{3D} = \sqrt{\frac{a_\text{con}\ps*{b_{\text{con},\,+} + b_{\text{con},\,-}}}{2}}
\end{equation}
By utilizing the procedure outlined in Section \ref{3D mechanical deformation model}, we accumulate a dataset of \(a_\text{3D}\) as a function of ball radius \(r_b\), applied load \(P\), and reduced elastic modulus \(E_r\), and perform least-squares minimization [\verb|scipy.optimize.least_squares(method=`Trust Region Reflective')| \cite{Virtanen2020}] on a function that mirrors the Hertzian contact radius expression:
\begin{equation}\label{eq:a3DModelFunction}
     a_\text{3D} = \ps*{\frac{c_0 P r_b}{E_r}}^{c_1}\,,
\end{equation}
where \(c_0\) and \(c_1\) are fitting constants to be found. Assuming a ball with the properties stated in the nomenclature Table \ref{table:NomenclatureBallBearings}, we find the following function to best fit the \(a_\text{3D}\) dataset; the parameter ranges used for the load \(P\), ball radius \(r_b\), and reduced elastic modulus \(E_r\) are also identical to those stated in the nomenclature table.
\begin{equation}\label{eq:a3DFittingFunction} 
     a_\text{3D} = \ps*{\frac{18.86 P r_b}{E_r}}^{0.3325}
\end{equation}

\section{Results \& Discussion}\label{Ball Bearing Results}
\begin{table}[H]
\centering
\begin{tabular}{||c | c ||} 
 \hline\hline
 \(H_\text{ball} = \SI{10}{GPa}\) & ball hardness \\ \hline
\(\nu = 0.27\) & Poisson's ratio \\ \hline
\(k_b = \SI{15.05}{\text{W}/(\text{m K})}\) & ball thermal conductivity \\ \hline
\(\varphi_b = 15^\circ\) (for 3D model) & bearing contact angle \\ \hline
\(k_f = \SI{24.2}{\text{W}/(\text{m K})}\) & thermal conductivity of flat and race \\ \hline
\(\rho_l = \SI{2200}{\text{kg}/\text{m}^3}\) &  lubricant density \\ \hline
\(\gamma_l = \SI{0.032}{\text{kg}/\text{s}^2}\) & surface tension of liquid on solid \\ \hline
\(\alpha_l = 1.5 \times 10^{-20}\,\text{J}\) & surface interaction strength \\ \hline
\(g = \SI{9.81}{\text{m}/\text{s}^2}\) & gravitational acceleration (towards flat/race) \\ \hline
\(k_l = \SI{0.16}{\text{W}/(\text{m K})}\) & thermal conductivity of lubricant \\ \hline
\(\frac{\text{lub. vol.}}{\text{ball vol.}}\) logarithmic space: \(5\times 10^{-6}\) - \(1 \times 10^{-2}\) & lubricant volume/ball volume range \\ \hline
\(r_b\) linear space: \SI{0.005}{m} - \SI{0.015}{m} & ball radius range \\ \hline
\(P\) logarithmic space: \SI{1}{N} - \SI{500}{N} & applied load range \\ \hline
\(E_r\) linear space: \SI{125}{GPa} - \SI{150}{GPa} & reduced elastic modulus range \\ \hline
\(a_\text{H}\): \(2.9\times 10^{-5}\) m - \(3.6\times 10^{-4}\) m & resulting Hertzian contact radius range \\\hline
\(a_\text{2D}\): \(2.9\times 10^{-5}\) m - \(3.6\times 10^{-4}\) m & resulting 2D multiphysics contact radius range \\\hline
\(a_\text{3D}\): \(8.7\times 10^{-5}\) m - \(1.1\times 10^{-3}\) m & resulting 3D multiphysics contact radius range \\[0.5ex]
 \hline\hline
\end{tabular}
\caption{Nomenclature and input parameters}
\label{table:NomenclatureBallBearings}
\end{table}

% \(\ps*{\frac{1 - \nu_\text{ball}^2}{E_\text{ball}} + \frac{1 - \nu_\text{flat}^2}{E_\text{flat}}}^{-1}\) 
% \item distance from bearing ring center to ball center \(r_\text{bearing} = \SI{0.3175}{m}\) (for 3D model)
% \item raceway radius \(r_r = 1.05 r_b\) (for 3D model)

\begin{table}[H]
\centering
\begin{tabular}{|| c | c ||} 
 \hline\hline
 \multirow{2}{0.95\linewidth}{number of \(\ps*{V_l, r_b, E_r}\) combinations for which training conductance data is generated for Yovanovich model} & 12500 \\
 & \\\hline
 \multirow{2}{0.95\linewidth}{number of \(\ps*{V_l, r_b, E_r}\) combinations for which training conductance data is generated for 2D and 3D multiphysics models} & 225 \\
 & \\\hline
 \multirow{2}{0.95\linewidth}{azimuthal angles used in solving for meniscus half geometry in 3D model} & 21 \\
 & \\\hline
 \multirow{2}{0.95\linewidth}{azimuthal angles used in solving for conductance contributions for meniscus half in 3D model} & 6 \\
 & \\[0.5ex]
 \hline\hline
\end{tabular}
\caption{Computational settings}
\label{table:ComputationalSettingsBallBearings}
\end{table}

With certain system inputs set as constant and others with specified ranges, we generated datasets of the thermal conductance ratio for each of the Yovanovich, 2D multiphysics, and 3D multiphysics models, which were then used to discover optimal fitting constants for the aforementioned correlation structures. Thus, correlations are presented here for each of the three lubricated ball bearing models and compare the results. In order to analyze the conductance ratio behavior of each model, we graph how the conductance ratio is affected both for varying load and for varying lubricant volume in the figures that follow. Additionally, we graph the means of the analytical correlation and the physical model outputs that the correlation fits are based on, to assess how closely the correlation reflects the underlying multiphysics model on average. However, averaging over the ball radii, loads, and reduced elastic moduli obfuscates the underlying distribution of the conductance ratios, so we also include a box plot alongside each graph of means to illustrate the spread of the data around the median at each normalized lubricant volume. 

Let us begin by discussing the structure of the correlations and their dimensionality. We reiterate that the thermal conductance ratio is expressed as
\begin{equation}\label{eq:ConductanceRatioGeneral}
     \text{conductance ratio} = \frac{G_\text{total}}{G_\text{dry}}\,,
\end{equation}
in which 
\begin{equation}\label{eq:ConductanceRatioYovRestate}
     \frac{G_\text{total, Yov.}}{G_\text{dry, Yov.}} = \frac{\pi k_l r_b}{2 a_\text{H} k_r} \bs*{I_\text{fit, Yov.}\ps*{V_l;\,r_b, a_\text{H}} - I_\text{fit, Yov.} \ps*{V_\text{min};\,r_b, a_\text{H}}} + 1
\end{equation}
for the Yovanovich model (as originally introduced in Eq. \eqref{eq:ConductanceRatioYovanovich}), and
\begin{equation}\label{eq:ConductanceRatio2Dor3D}
     \frac{G_\text{total, 2D or 3D}}{G_\text{dry, 2D or 3D}} = \frac{\pi k_l r_b}{2 a_\text{2D or 3D} k_r} I_\text{fit, 2D or 3D}\ps*{V_l; r_b, a_\text{2D or 3D}} + 1
\end{equation}
for the 2D or 3D multiphysics models, where \(I_\text{fit}\) represents a fitting function that can map lubricant volume \(V_l\) to the total lubricant conductance integral of Eq. \eqref{eq:ConductanceIntegral} divided by ball radius for the Yovanovich model, or its analogous value for the 2D and 3D multiphysics models which is determined by our numerical methodology. Through numerical inspection of how Eq. \eqref{eq:YovanovichIntegralResultIndefinite} indirectly depends on the the lubricant volume, we discovered that a suitable form for \(I_\text{fit}\) is a ``slope'' multiplied onto the log of normalized lubricant volume \(V_l\) plus an ``intercept''. The ``slope'' and ``intercept'' are each a polynomial in the natural log of normalized contact radius \(a\). Writing these polynomials as finite sums, we can express the form of \(I_\text{fit}\) generically, where the polynomials may be of arbitrarily high order.
\begin{equation}\label{eq:IFitGenerating}
     I_\text{fit, linear}\ps*{V_l; r_b, a} = \ps*{\sum_{i=0}^n c_i \ln^i \frac{a}{r_b}} \ln \ps*{\frac{V_l}{\frac{4}{3}\pi r_b^3}} + \sum_{i=n+1}^m c_i \ln^i \frac{a}{r_b}
\end{equation}
As we introduced in Section \ref{Correlation development for 2D and 3D multiphysics models}, we found that setting \(\ps*{n = 1,\, m = 3}\) for a total of 4 fitting constants yielded sufficient accuracy to the underlying numerical models. Again, it is imperative that \(r_b I_\text{fit}\) has the same dimensionality as the lubricant conductance integral Eq. \eqref{eq:ConductanceIntegral} with units of length, as the integral physically implies that the lubricant conductance should be proportional to the ball radius \(r_b\). It is reasonable to assert that this physical implication should extend to the 2D and 3D multiphysics models, because the lubricant meniscii of all the models only differ in having flat or curved edges at their maximum radial extents, and in having azimuthal symmetry or a lack thereof. As a consequence of the dimensionality in length arising from \(r_b I_\text{fit}\), the terms within the \(I_\text{fit}\) correlations themselves may be non-dimensionalized with respect to the ball radius or the ball volume.

In terms of the utility of the following \(I_\text{fit}\) model correlations, they apply for the specific ball radii, applied loads, normalized lubricant volumes, and lubricant properties listed in the nomenclature Table \ref{table:NomenclatureBallBearings}. The contact radius ranges that result from the mechanical deformation calculations of sections \ref{Methods for ball-on-flat model} and \ref{Methods for ball-on-races model} are also listed in Table \ref{table:NomenclatureBallBearings}. Note that the correlations are independent of the thermal conductivities because the conductivities are part of the coefficient to \(I_\text{fit}\) in the conductance ratio equations \eqref{eq:ConductanceRatioYovRestate}-\eqref{eq:ConductanceRatio2Dor3D}. Extrapolating conductance ratios for inputs outside of the listed parameter ranges is likely most inaccurate for normalized lubricant volumes outside of the listed training data, as the correlations should be particularly sensitive to the ball-on-surface maximum lubricant meniscus volume which is a consequence of the geometric constraints.
\begin{align}\label{eq:YovModelCorrelationResult}
    \begin{split}
    	I_\text{fit, Yov.}\ps*{V_l;\,r_b, a_\text{H}} & = \ps*{-8.58\times 10^{-5}\cdot \ln{\frac{a_\text{H}}{r_b}} + 4.97\times 10^{-1}}\ln\ps*{\frac{V_l}{\frac{4}{3} \pi r_b^3}}\\
        & -1.15\times 10^{-1}\cdot \ln{\frac{a_\text{H}}{r_b}} - 3.87
    \end{split}
\end{align}

\begin{figure}[H]
\centering
\begin{subfigure}{.49\textwidth}
    \centering
    \includegraphics[width=1\linewidth]{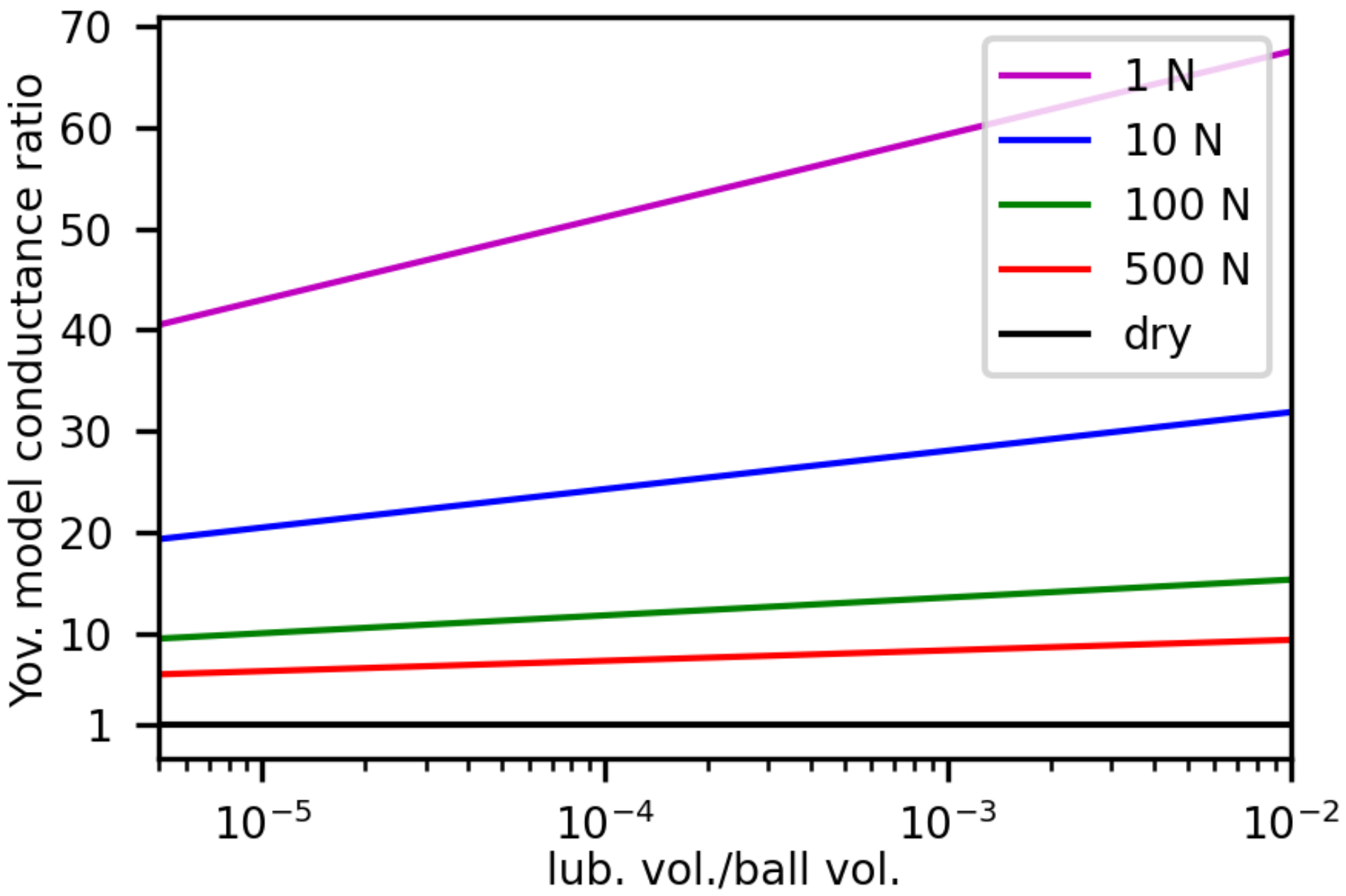}
    \caption{cond. ratio vs. normalized lubricant vol.}
    \label{fig:YovCondRatioVol}
\end{subfigure}
\hfill
\begin{subfigure}{.49\textwidth}
    \centering
    \includegraphics[width=1\linewidth]{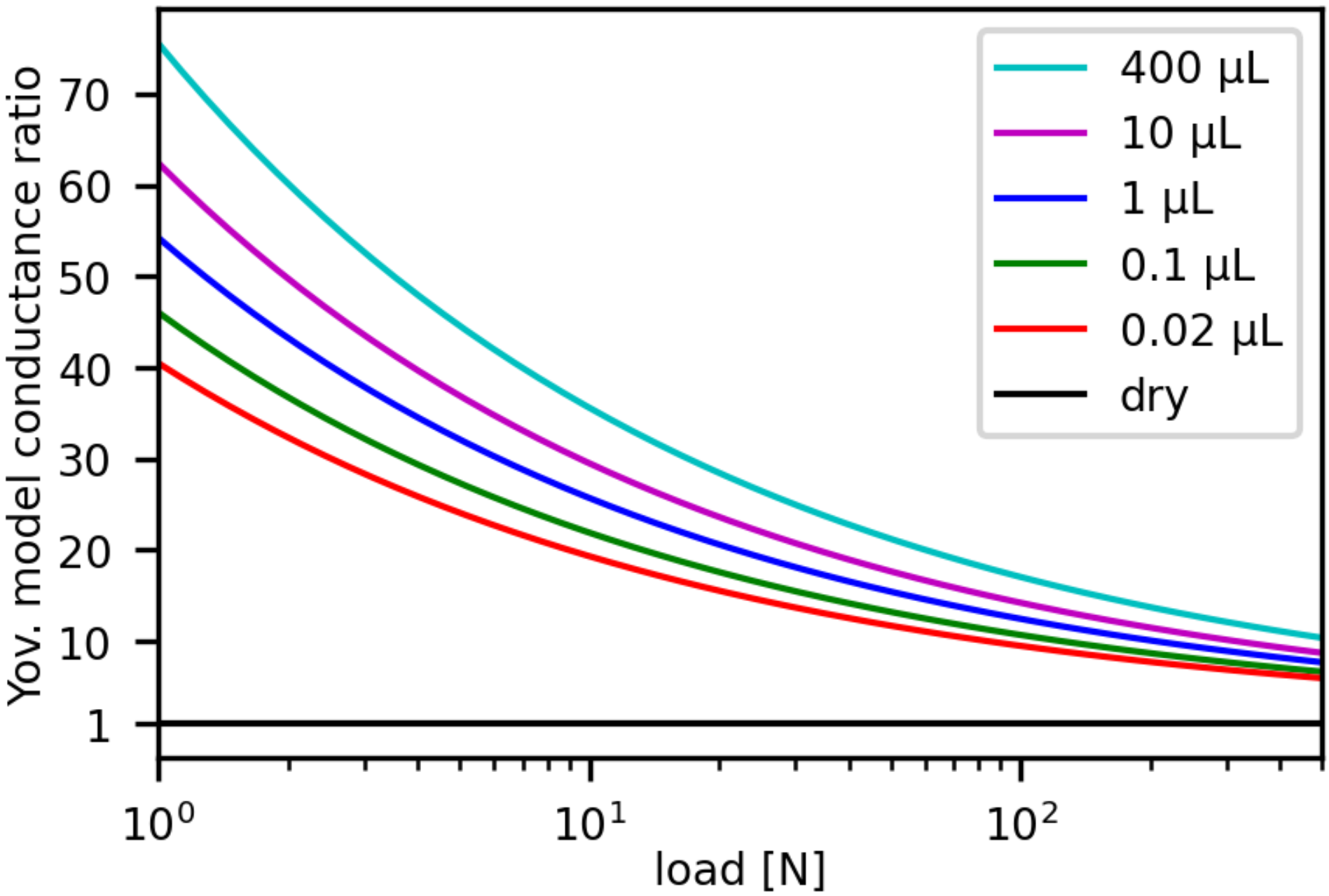}
    \caption{cond. ratio vs. applied load}
    \label{fig:YovCondRatioLoad}
\end{subfigure}
\caption{\(G_\text{total, Yov.}/G_\text{dry}\) plotted for the inputs: \(r_b = \SI{0.01}{m}\), \(E_r = \SI{137.5}{GPa}\), \(k_\text{lub}/\ps*{k_\text{ball}^{-1} + k_\text{flat}^{-1}}^{-1} = 0.055\)}
\label{fig:YovCondRatios}
\end{figure}

Eq. \eqref{eq:YovModelCorrelationResult} provides the 4-parameter fit determined for the conductance of the ball-on-flat system with varying Hertzian contact radius and lubricant meniscus volume according to the method of Yovanovich in subsection \ref{Correlation development using Yovanovich's numerical model}, and the equation below provides the 4-parameter fit determined for the conductance of the ball-on-flat system with varying contact radius and lubricant meniscus volume according to the 2D multiphysics methods of section \ref{Methods for ball-on-flat model}. Note that while the graph of the means shows the mean values of the physical model and the analytical correlation, the mean error is defined as the average of the errors between the underlying physical model data points and the correlation fit points, which we depict the respective spreads of in the corresponding box plot. The box plot shows boxes with whiskers for the physical model data and analytical correlation data at each normalized lubricant volume, where the physical model boxes (colored black) are shifted slightly left and the analytical correlation boxes (colored blue) are shifted slightly right so as to not visually overlap each other. Each box extends from the first quartile to the third quartile of the data with the red line at the median, and the whiskers extend from a box to the farthest point that lies within 1.5 times the interquartile range from that box. Below, we see very good agreement between both the physical model and analytical correlation mean values as well as their underlying distributions for the Yovanovich model.

\begin{figure}[H]
\centering
\begin{subfigure}{.49\textwidth}
    \centering
    \includegraphics[width=1\linewidth]{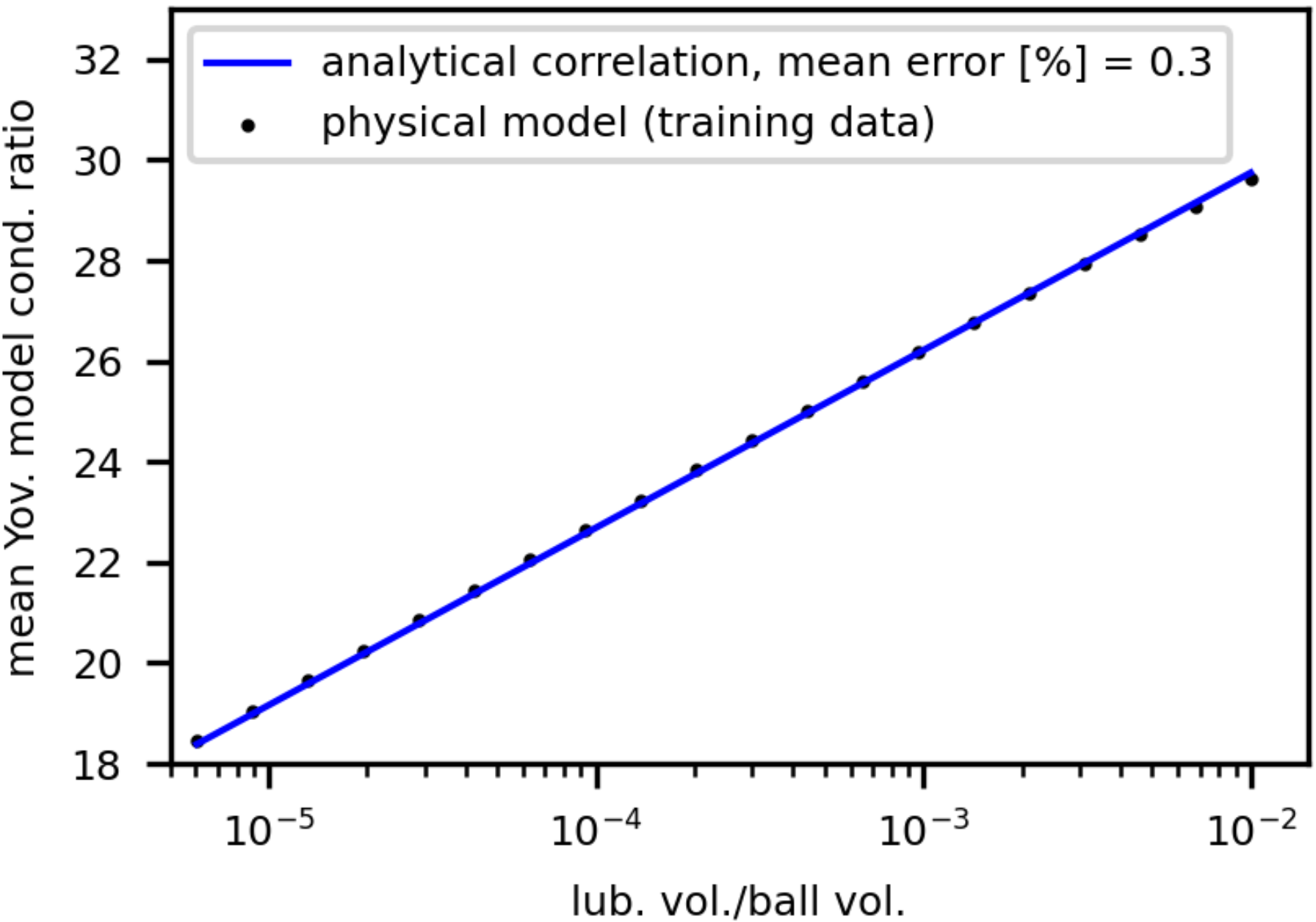}
    \caption{means, averaged across all \(r_b,\,P,\,E_r\)}
    \label{fig:YovMeans}
\end{subfigure}
\hfill
\begin{subfigure}{.49\textwidth}
    \centering
    \includegraphics[width=1\linewidth]{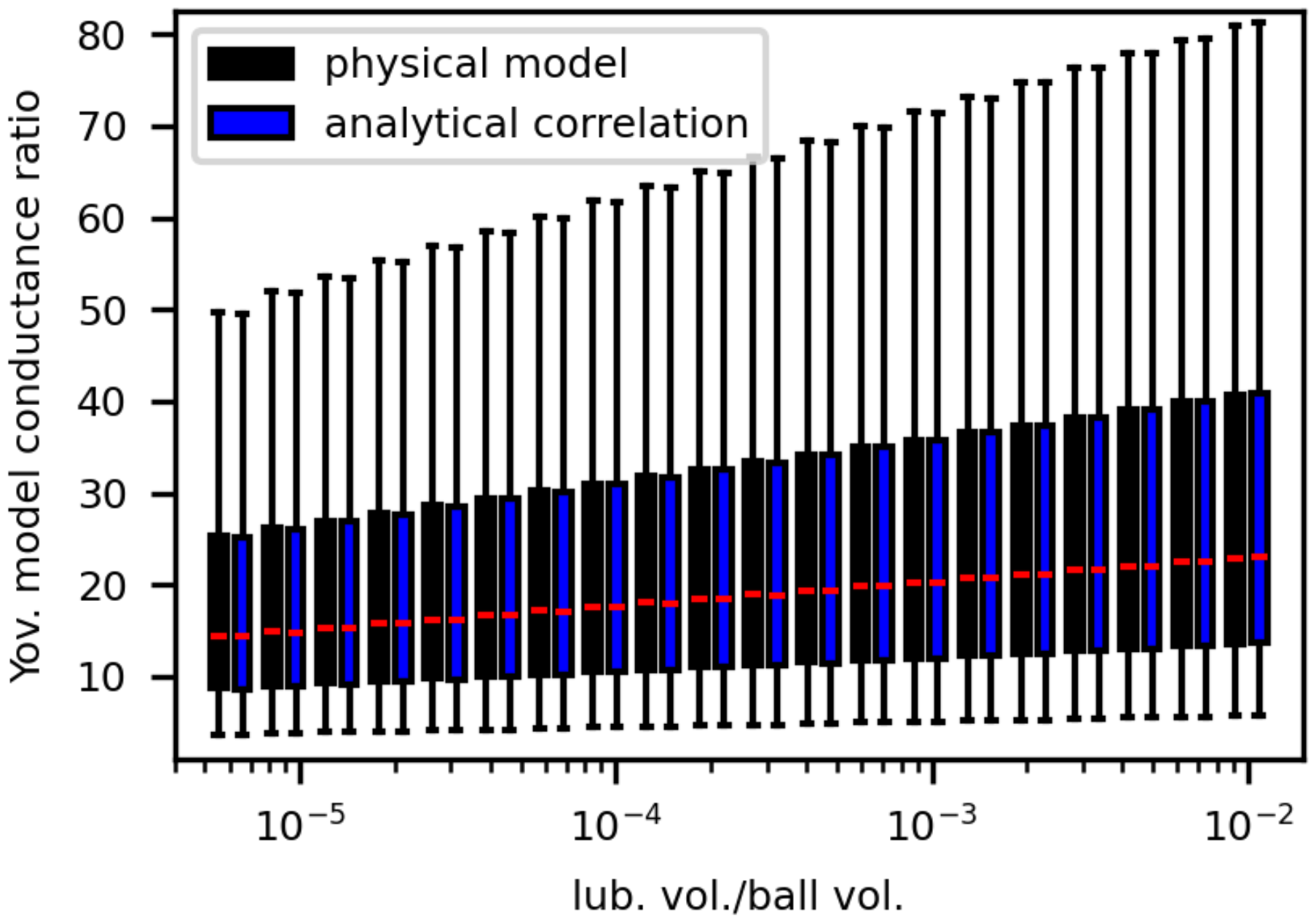}
    \caption{underlying distribution in \(r_b,\,P,\,E_r\)}
    \label{fig:YovMedians}
\end{subfigure}
\caption{Comparison of Yovanovich analytical correlation and physical model calculations as function of normalized lubricant volume}
\label{fig:YovMeansMedians}
\end{figure}

Next, we compare the thermal conductance ratios calculated by Yovanovich's ball-on-flat model to those of the 2D multiphysics ball-on-flat model. It follows that the case of a dry ball-on-flat has a conductance ratio equal to one. We first observe that for both of these models, decreasing the applied load or increasing the lubricant volume increases the conductance ratio. The main contrast in the models' results is that Yovanovich's model predicts a significantly higher conductance ratio, approximately 5 to 10 times that of the conductance ratio generated by the 2D multiphysics model of the present work, depending on the lubricant volume and load. This outcome can be attributed to Yovanovich's lubricant meniscus covering a larger contact area, and therefore yielding a larger heat transfer pathway for the same lubricant volume given in both models. The difference in contact area is due to how, in Yovanovich's meniscus model the meniscus is treated as a vertical ``wall'', instead of having curvature and conforming closer to the ball and flat surfaces caused by a balance of meniscus forces as per our multiphysics model, leading to Yovanovich's meniscus having larger lubricated contact area. 

\begin{align}\label{eq:2DModelCorrelationResult}
    \begin{split}
    I_\text{fit, 2D}\ps*{V_l;\,r_b, a_\text{2D}} & = \ps*{-1.19\times 10^{-2}\cdot \ln{\frac{a_\text{2D}}{r_b}} + 2.26\times 10^{-1}}\ln\ps*{\frac{V_l}{\frac{4}{3} \pi r_b^3}}\\
    & - 3.45\times 10^{-1}\cdot \ln{\frac{a_\text{2D}}{r_b}} + 2.29
    \end{split}
\end{align}

\begin{figure}[H]
\centering
\begin{subfigure}{.49\textwidth}
    \centering
    \includegraphics[width=1\linewidth]{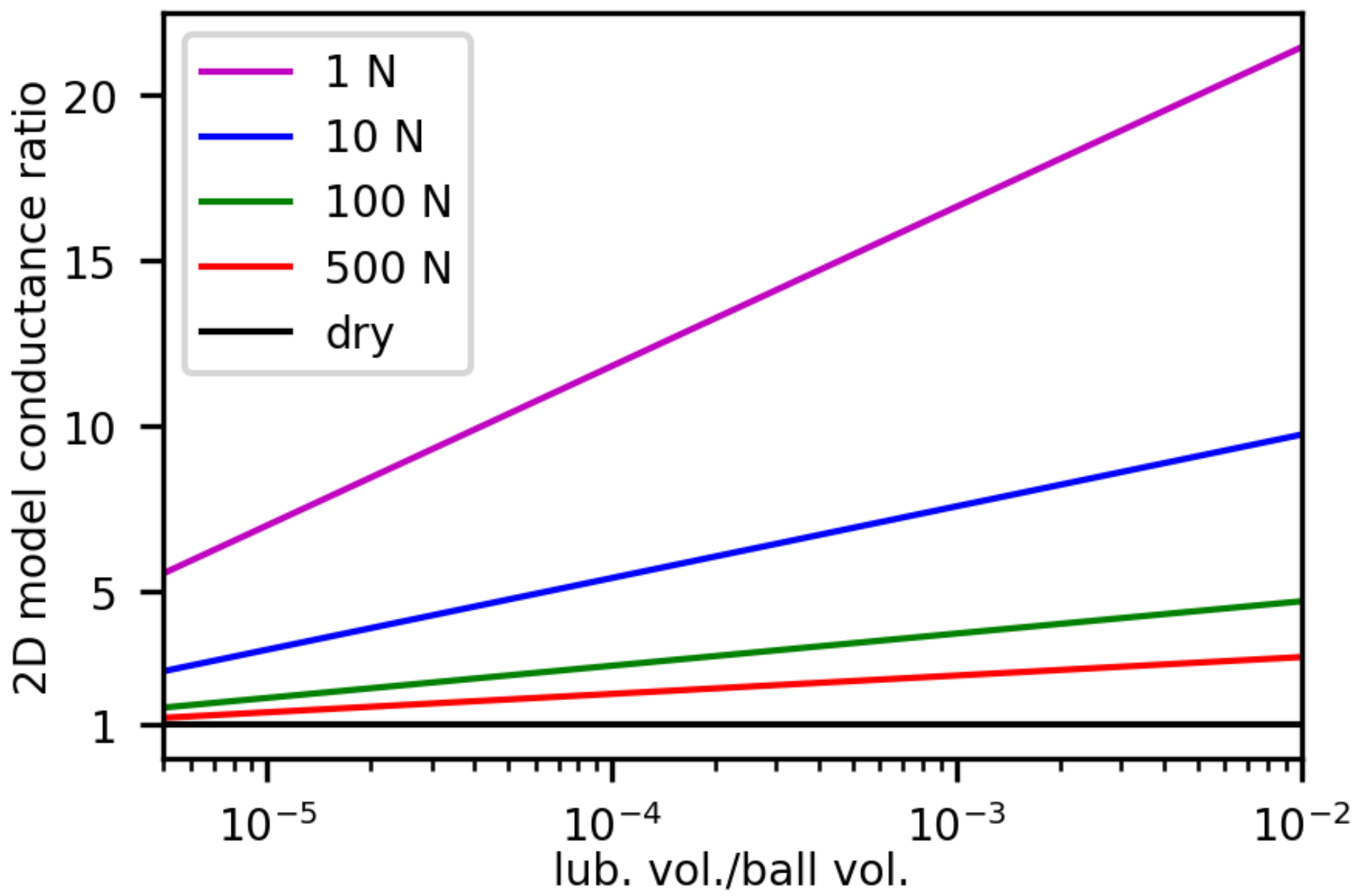}
    \caption{cond. ratio vs. normalized lubricant vol.}
    \label{fig:2DCondRatioVol}
\end{subfigure}
\hfill
\begin{subfigure}{.49\textwidth}
    \centering
    \includegraphics[width=1\linewidth]{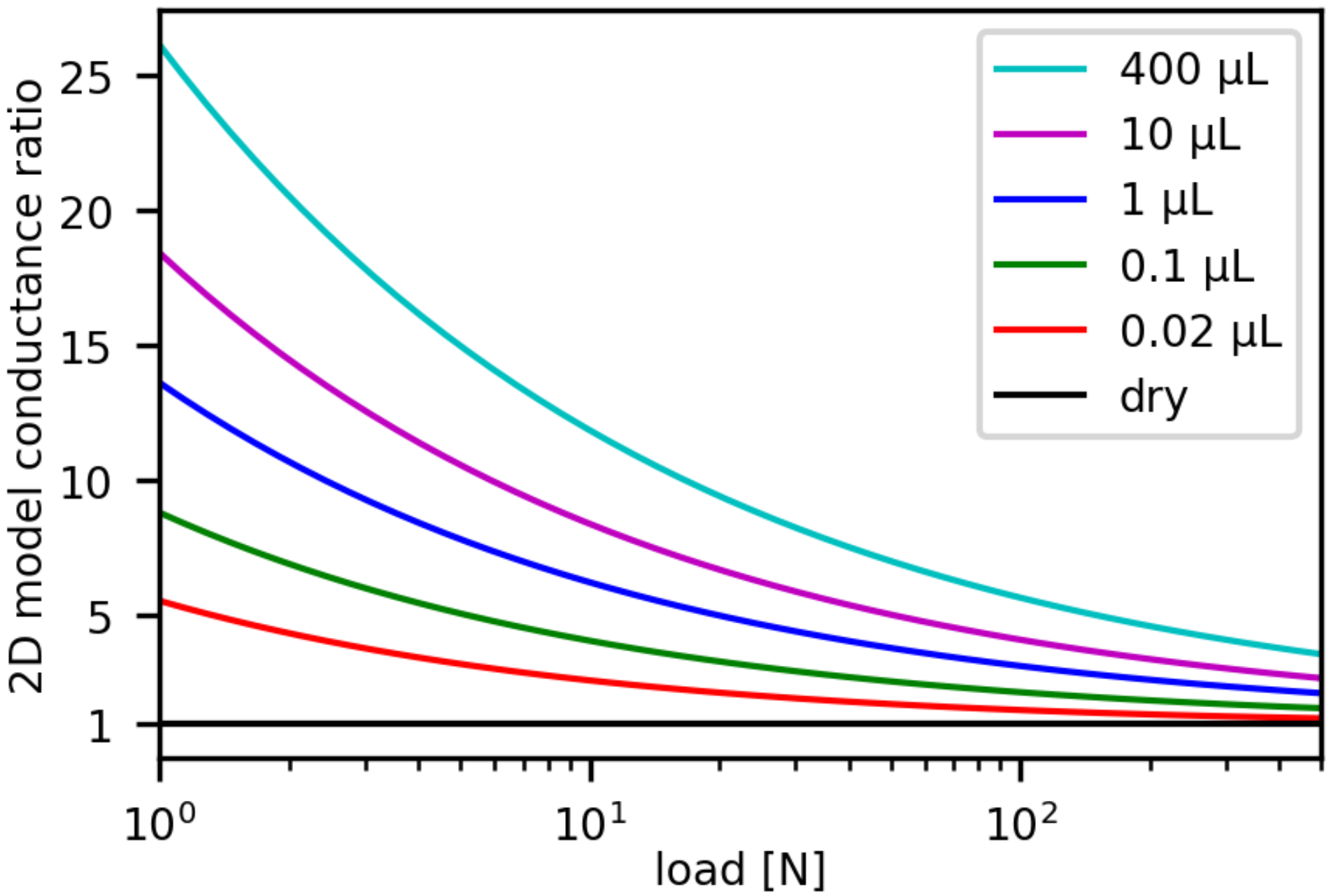}
    \caption{cond. ratio vs. applied load}
    \label{fig:2DCondRatioLoad}
\end{subfigure}
\caption{\(G_\text{total, 2D}/G_\text{dry, 2D}\) plotted for the inputs: \(r_b = \SI{0.01}{m}\), \(E_r = \SI{137.5}{GPa}\), \(k_\text{lub}/\ps*{k_\text{ball}^{-1} + k_\text{flat}^{-1}}^{-1} = 0.055\)}
\label{fig:2DCondRatios}
\end{figure}

\begin{figure}[H]
\centering
\begin{subfigure}{.49\textwidth}
    \centering
    \includegraphics[width=1\linewidth]{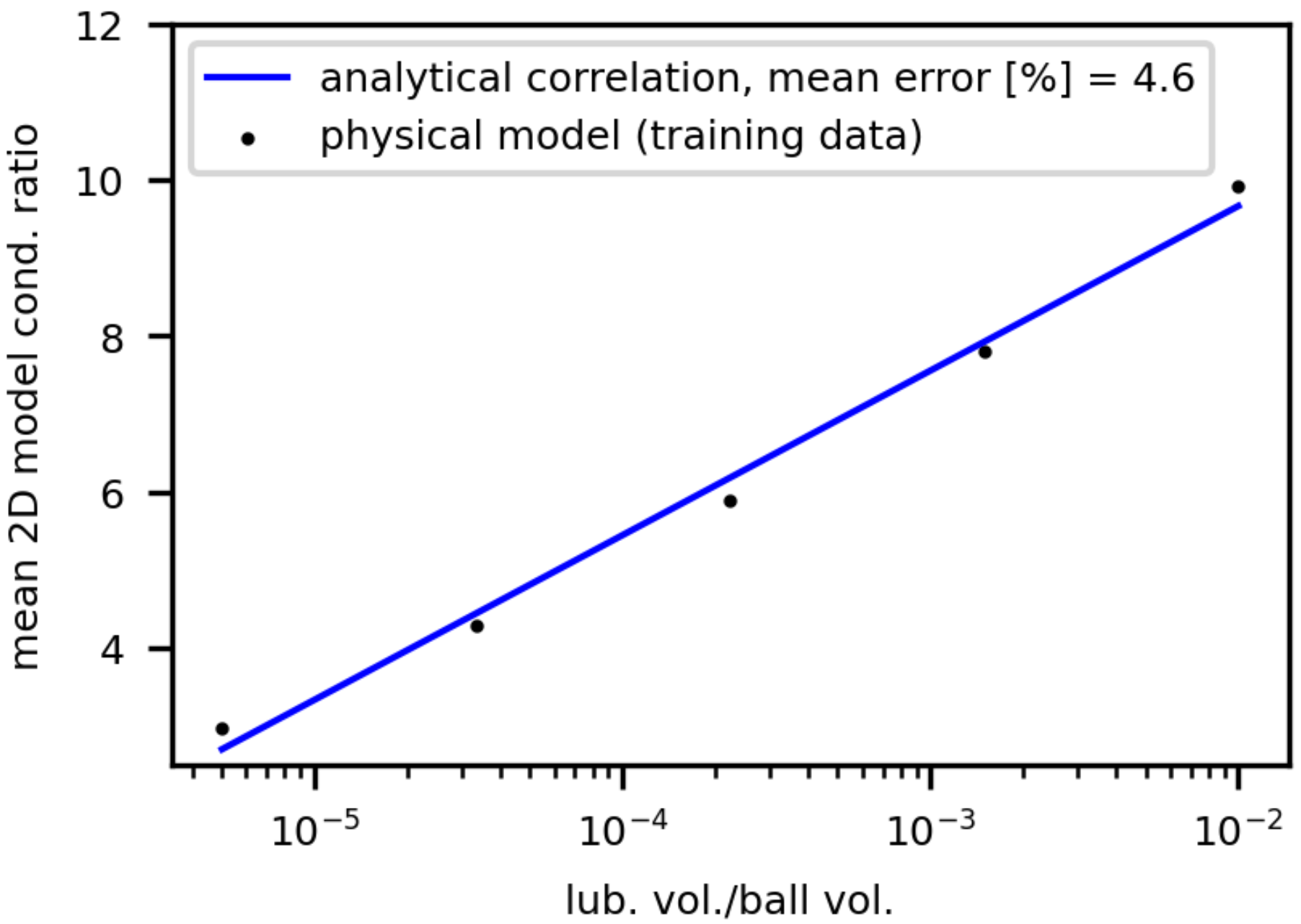}
    \caption{means, averaged across all \(r_b,\,P,\,E_r\)}
    \label{fig:2DMeans}
\end{subfigure}
\hfill
\begin{subfigure}{.49\textwidth}
    \centering
    \includegraphics[width=1\linewidth]{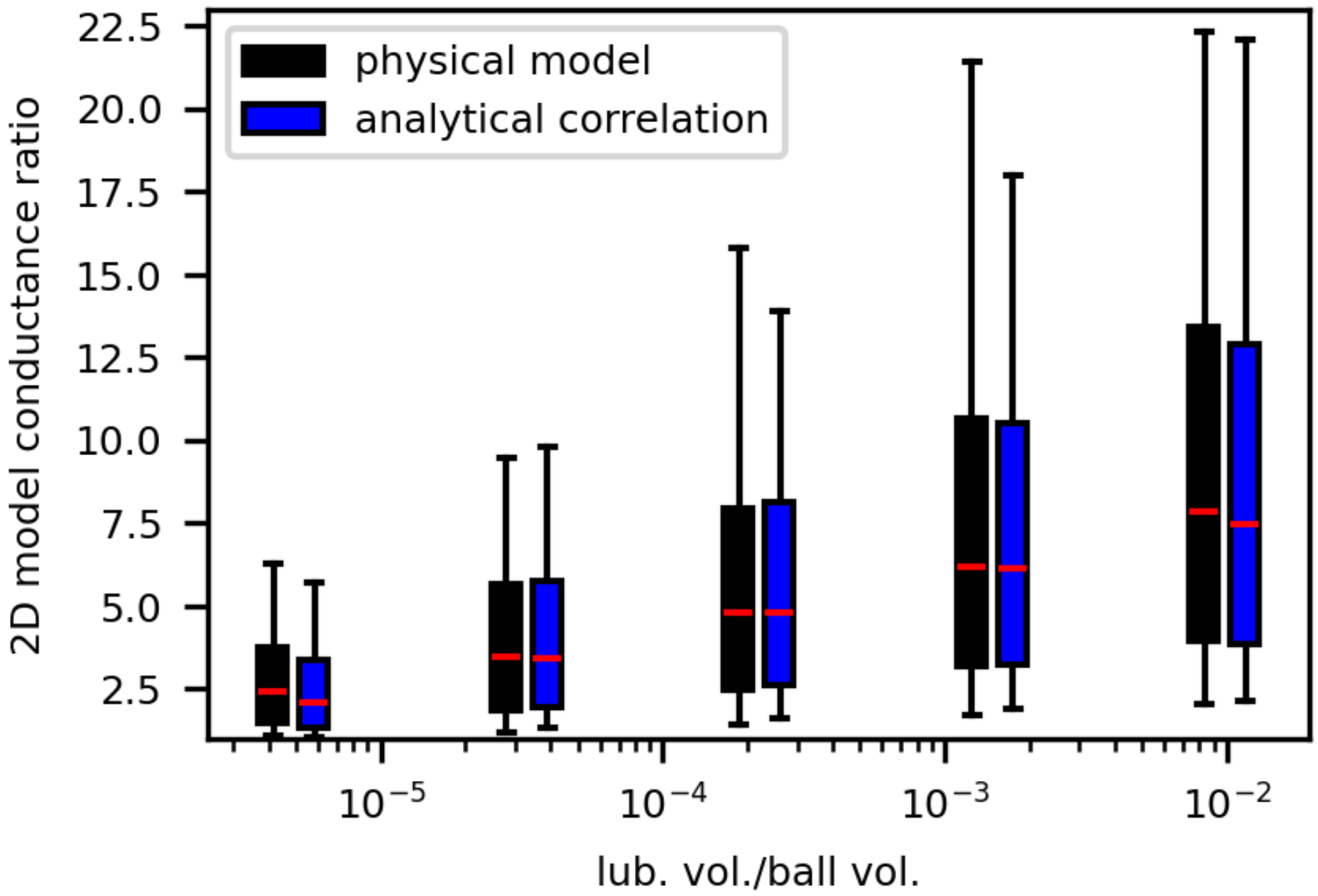}
    \caption{underlying distribution in \(r_b,\,P,\,E_r\)}
    \label{fig:2DMedians}
\end{subfigure}
\caption{Comparison of 2D multiphysics-based analytical correlation and physical model calculations as function of normalized lubricant volume}
\label{fig:2DMeansMedians}
\end{figure}

\begin{figure}[H]
\centering
\begin{subfigure}{.49\textwidth}
    \centering
    \includegraphics[width=1\linewidth]{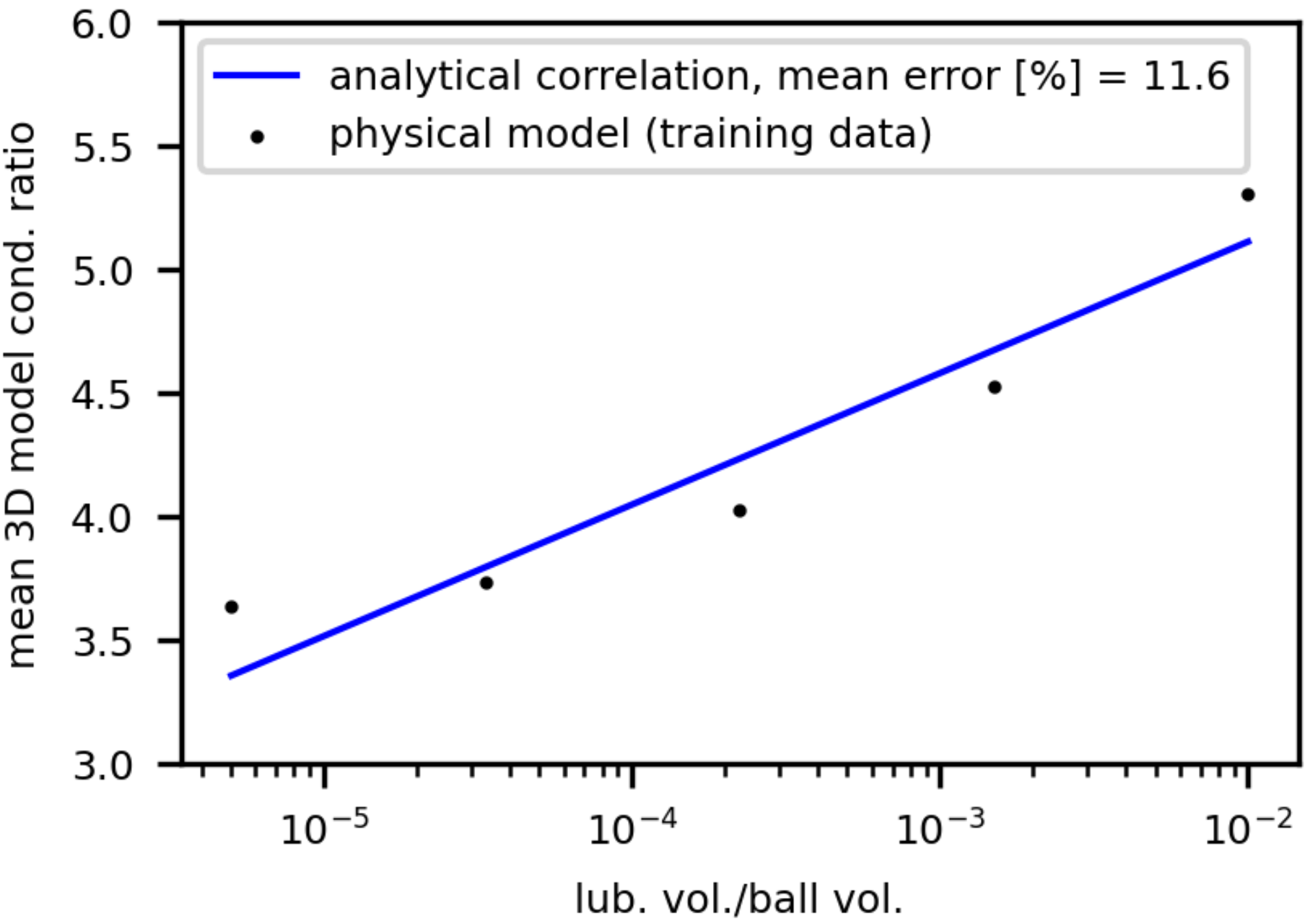}
    \caption{means, averaged across all \(r_b,\,P,\,E_r\)}
    \label{fig:3DMeans}
\end{subfigure}
\hfill
\begin{subfigure}{.49\textwidth}
    \centering
    \includegraphics[width=1\linewidth]{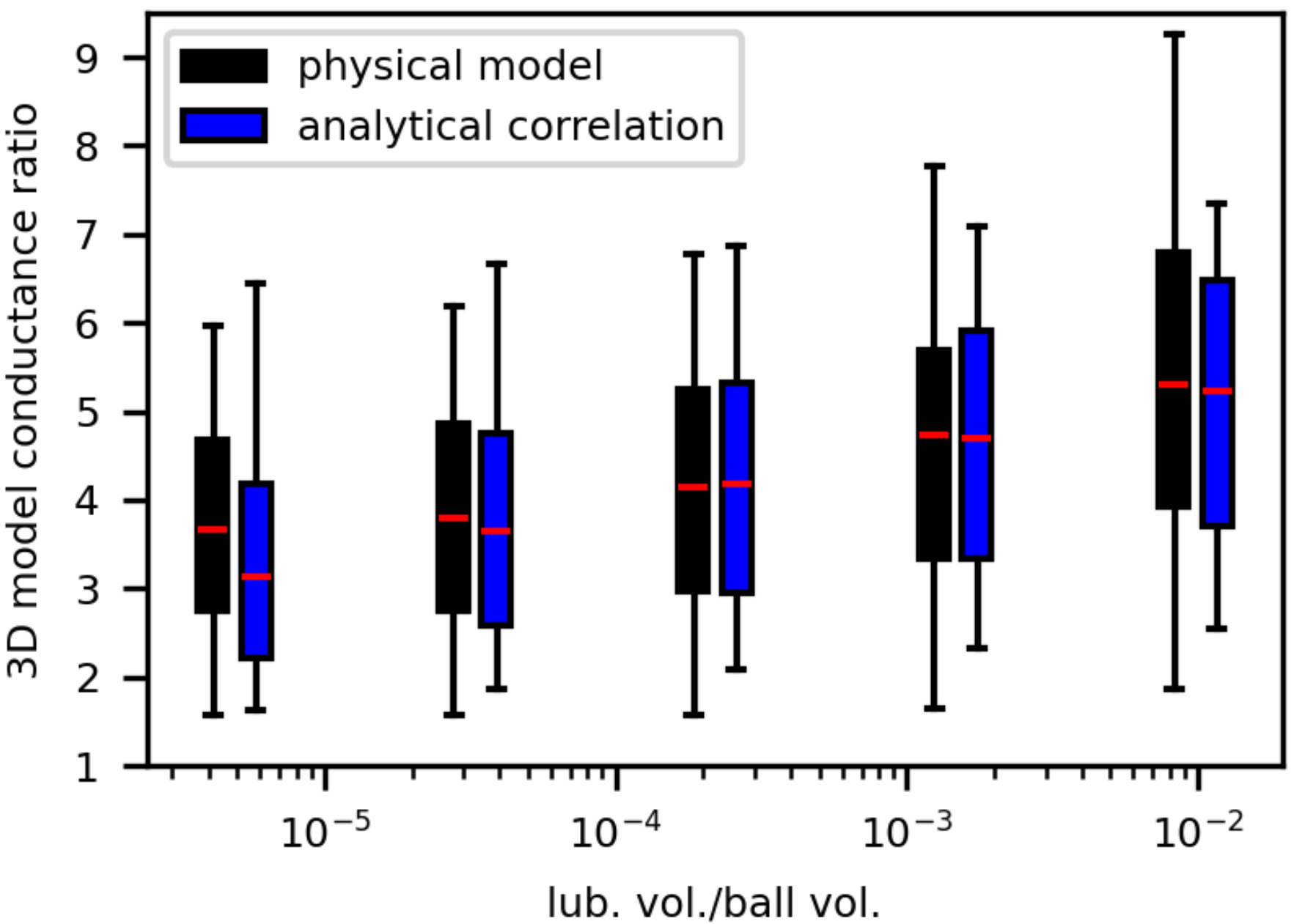}
    \caption{underlying distribution in \(r_b,\,P,\,E_r\)}
    \label{fig:3DMedians}
\end{subfigure}
\caption{Comparison of 3D multiphysics-based analytical correlation with 4 fitting constants and physical model calculations as function of normalized lubricant volume}
\label{fig:3DMeansMedians}
\end{figure}

Regarding the comparisons of the 2D and 3D 4-parameter correlations to their corresponding physical model datasets in Figs. \ref{fig:2DMeansMedians} and \ref{fig:3DMeansMedians}, we notice that the means of both correlations underpredict with respect to the physical model at the lubricant volume extremes, and overpredict with respect to the physical model near the center of the lubricant volume range; on the other hand, the medians of the correlation datasets also give underprediction at the lubricant volume extremes, but do not display significant overprediction near the center of the lubricant volume range. 

While the mean errors for the Yovanovich and 2D multiphysics correlations are both under 5\% (Figs. \ref{fig:YovMeans} and \ref{fig:2DMeans}), the mean error in the 4-parameter 3D multiphysics correlation reaches 11.6\% (Fig. \ref{fig:3DMeans}); the increased mean error with respect to the underlying multiphysics outputs for the 2D and 3D multiphysics models is likely due to the correlation structure inspired by Yovanovich's integral Eq. \eqref{eq:ConductanceIntegralEvaluated} inaccurately representing the total lubricant conductance in these models, which does not seem to be linear-log in the lubricant volume variable as it is in the Yovanovich model. To improve the accuracy of the 2D and 3D multiphysics models, it may be useful to test higher order polynomials for the slope and intercept terms in the correlation structure, or a polynomial in the log of lubricant volume term, and observe if the potentially higher accuracy merits the use of more fitting constants. To this end, we also try the following quadractic-log correlation structure with 6 fitting parameters, i.e., \(\ps*{n = 1,\, m = 3,\, N = 5}\), for the 3D multiphysics model:
\begin{align}\label{eq:IFitGenerating2}
    \begin{split}
    I_\text{fit, quadratic}\ps*{V_l; r_b, a} & = \ps*{\sum_{i=0}^n c_i \ln^i \frac{a}{r_b}} \ln^2 \ps*{\frac{V_l}{\frac{4}{3}\pi r_b^3}} + \ps*{\sum_{i=n+1}^m c_i \ln^i\frac{a}{r_b}} \ln \ps*{\frac{V_l}{\frac{4}{3}\pi r_b^3}}\\
    & + \sum_{i=m+1}^N c_i \ln^i \frac{a}{r_b}\,.
    \end{split}
\end{align}
Thus, the equations below provide the 4-parameter (\(I_\text{fit, 3D, 4}\)) and 6-parameter (\(I_\text{fit, 3D, 6}\)) fits determined for the conductance of the ball-on-race system with varying contact radius and lubricant meniscus volume according to the 3D multiphysics methods in section \ref{Methods for ball-on-races model}. 
\begin{align}\label{eq:3DModelCorrelationResult}
    \begin{split}
    I_\text{fit, 3D, 4}\ps*{V_l;\,r_b, a_\text{3D}} & = \ps*{2.18\times 10^{-1}\cdot \ln{\frac{a_\text{3D}}{r_b}} + 1.08}\ln\ps*{\frac{V_l}{\frac{4}{3} \pi r_b^3}}\\
    & + 3.27 \cdot \ln{\frac{a_\text{3D}}{r_b}} + 17.74
    \end{split}
\end{align}
\begin{figure}[H]
\centering
\begin{subfigure}{.49\textwidth}
    \centering
    \includegraphics[width=1\linewidth]{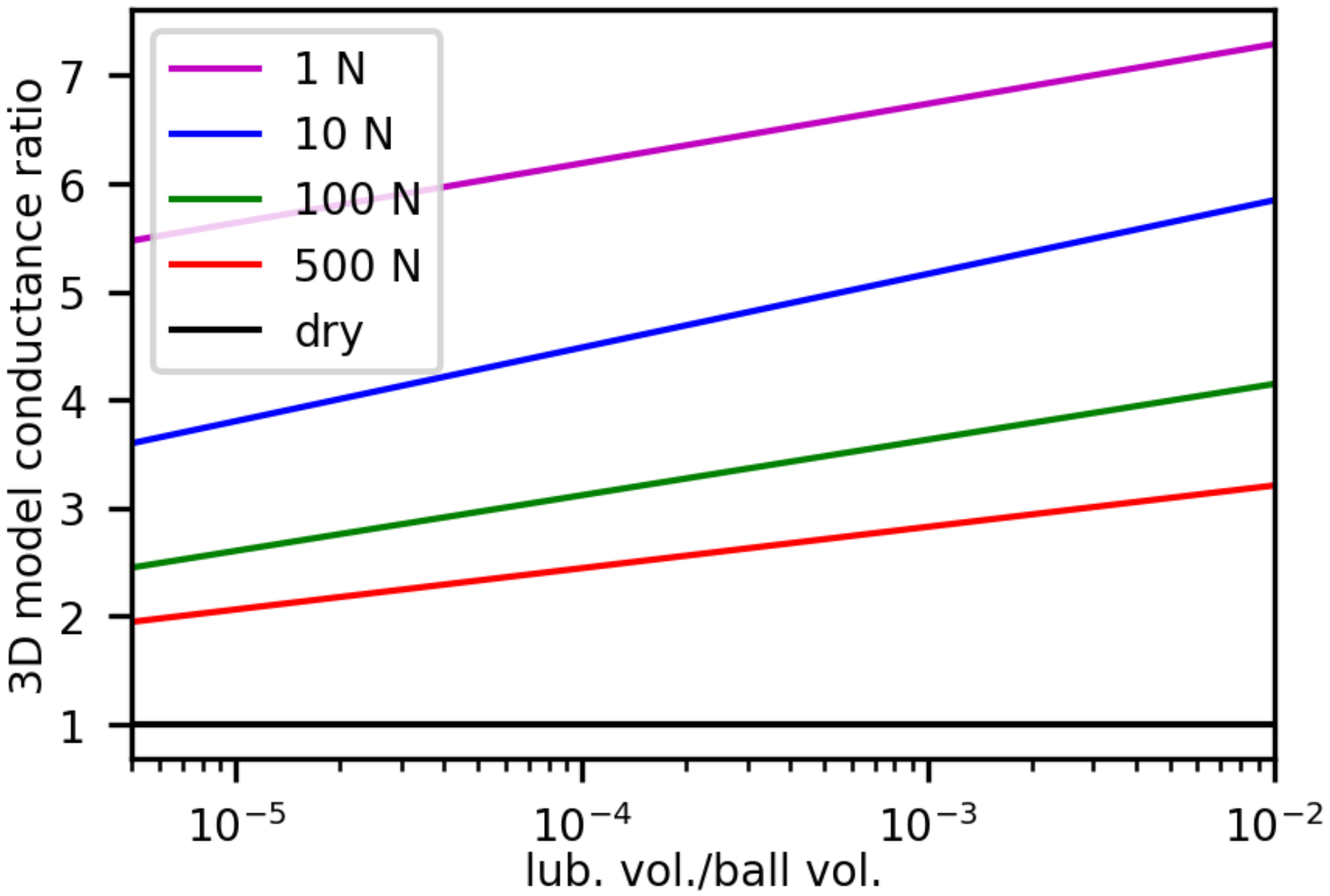}
    \caption{cond. ratio vs. normalized lubricant vol.}
    \label{fig:3DCondRatioVol}
\end{subfigure}
\hfill
\begin{subfigure}{.49\textwidth}
    \centering
    \includegraphics[width=1\linewidth]{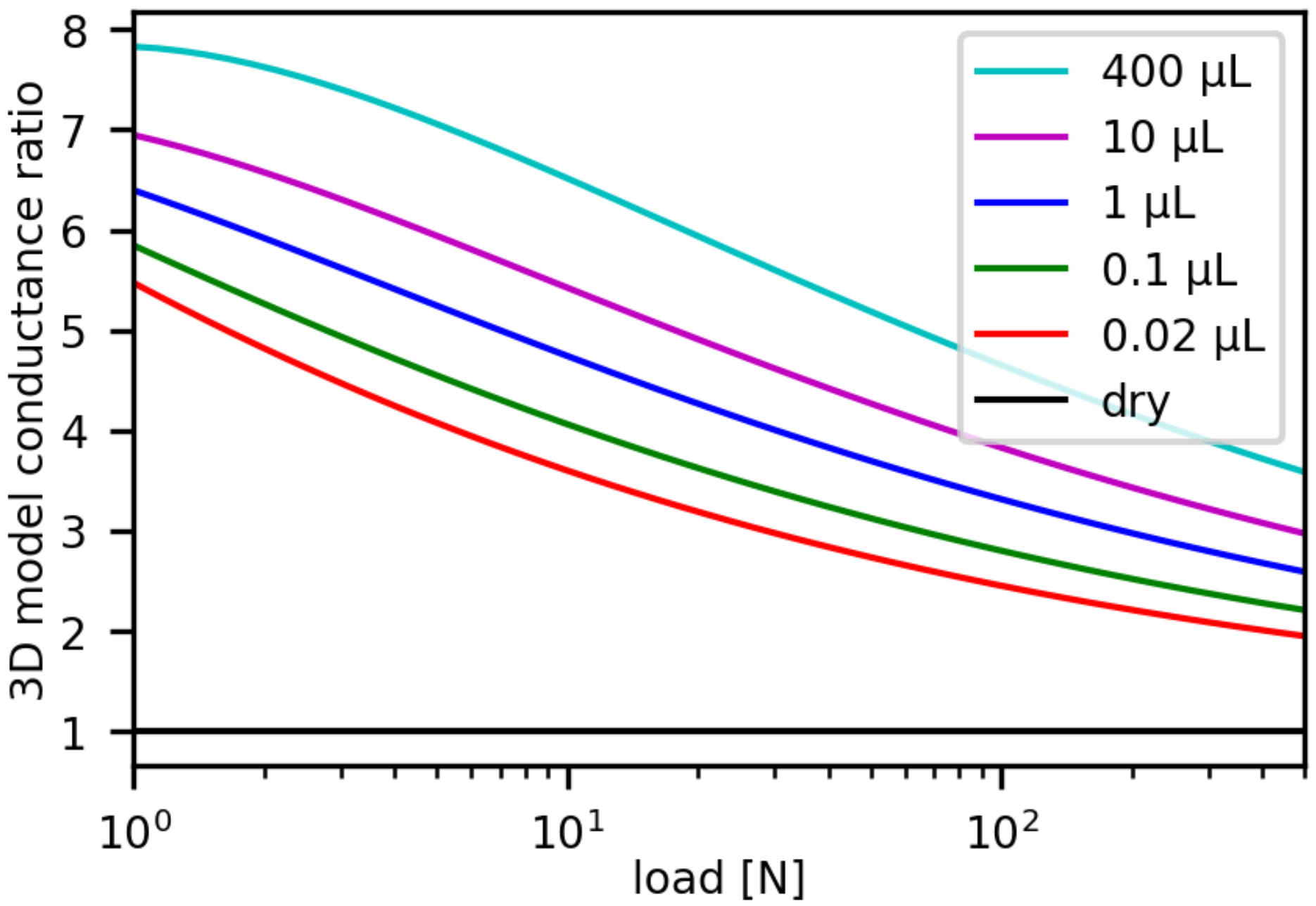}
    \caption{cond. ratio vs. applied load}
    \label{fig:3DCondRatioLoad}
\end{subfigure}
\caption{\(G_\text{total, 3D}/G_\text{dry, 3D}\) with 4 fitting constants plotted for the inputs: \(r_b = \SI{0.01}{m}\), \(E_r = \SI{137.5}{GPa}\), \(k_\text{lub}/\ps*{k_\text{ball}^{-1} + k_\text{race}^{-1}}^{-1} = 0.055\)}
\label{fig:3DCondRatios}
\end{figure}

Comparing the 2D and 3D multiphysics 4-parameter correlations, the maximum conductance ratio of the 3D model is approximately 5 times lower than that of the 2D model. And, at the extremes of highest lubricant volume or lowest load, the conductance ratio variation is significantly smaller in the 3D model versus in the 2D model for varying load or lubricant volume, respectively. This can be attributed to how the lubricant meniscus geometries for the 3D model across a range of volumes and loads are more similar to each other, than the lubricant geometries for the 2D model which feature more dramatic changes to the heat transfer pathway for varying volume and load.

\begin{figure}[H]
\centering
\begin{subfigure}{.49\textwidth}
    \centering
    \includegraphics[width=1\linewidth]{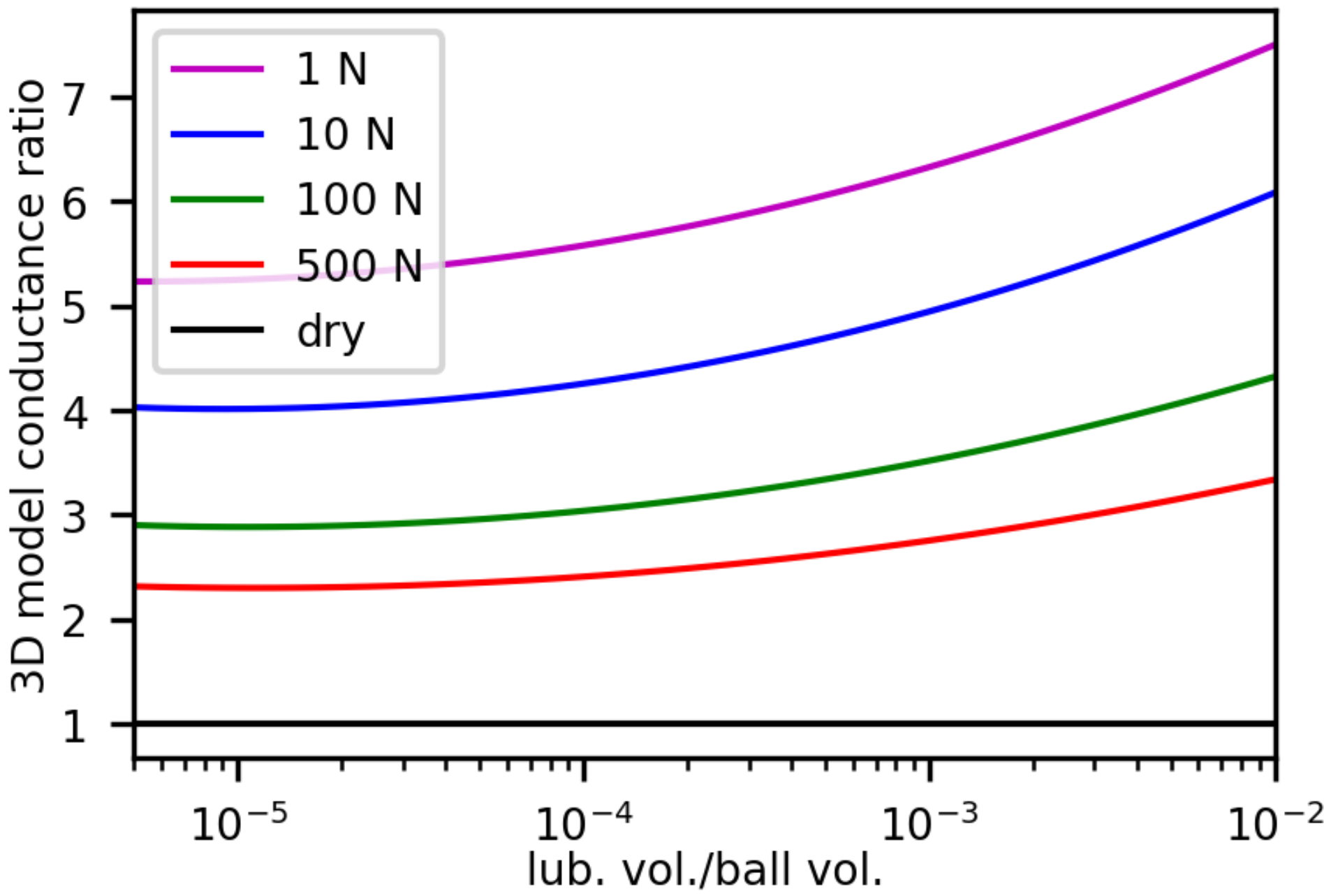}
    \caption{cond. ratio vs. normalized lubricant vol.}
    \label{fig:3DCondRatioVol2}
\end{subfigure}
\hfill
\begin{subfigure}{.49\textwidth}
    \centering
    \includegraphics[width=1\linewidth]{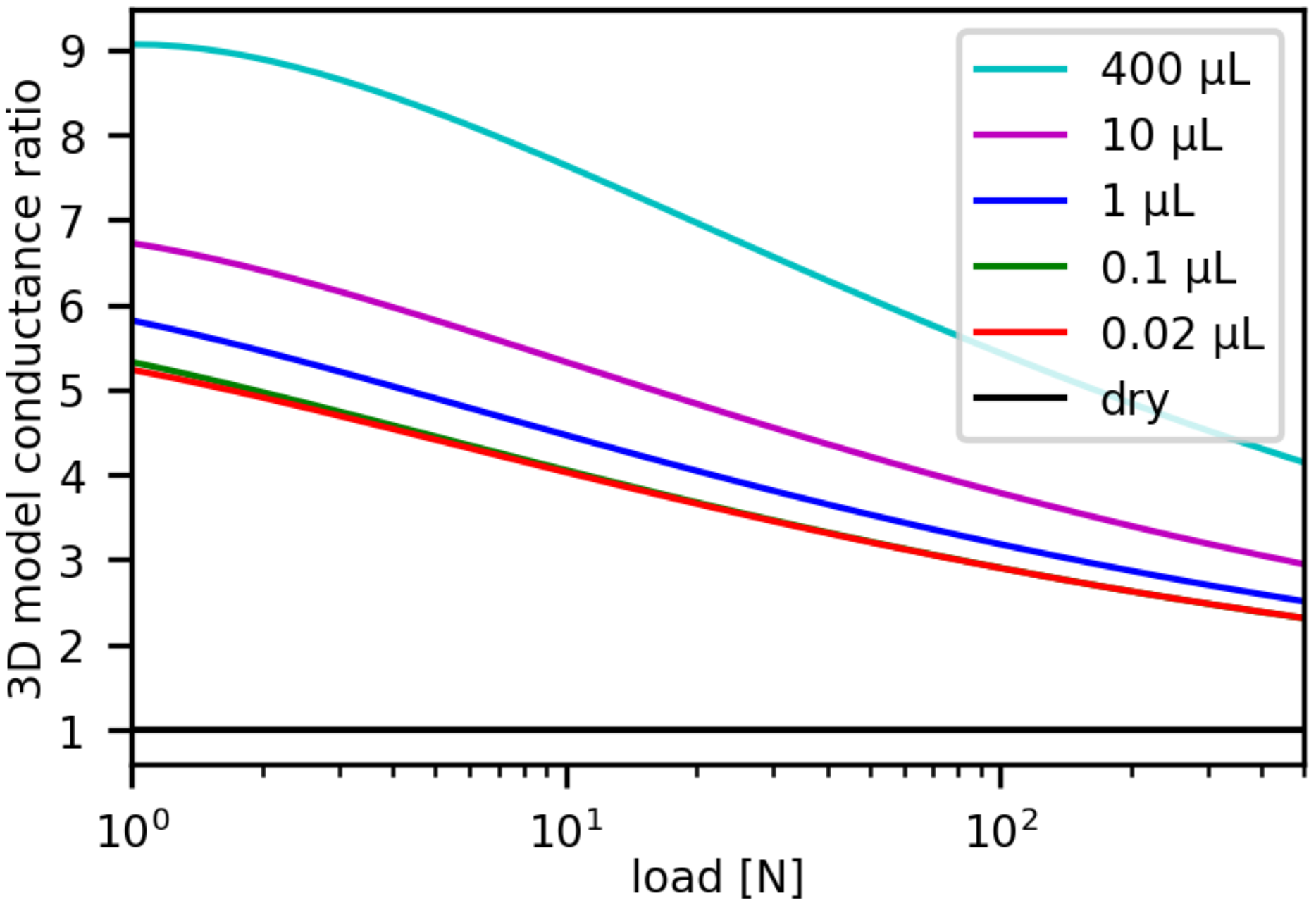}
    \caption{cond. ratio vs. applied load}
    \label{fig:3DCondRatioLoad2}
\end{subfigure}
\caption{\(G_\text{total, 3D}/G_\text{dry, 3D}\) with 6 fitting constants plotted for the inputs: \(r_b = \SI{0.01}{m}\), \(E_r = \SI{137.5}{GPa}\), \(k_\text{lub}/\ps*{k_\text{ball}^{-1} + k_\text{race}^{-1}}^{-1} = 0.055\)}
\label{fig:3DCondRatios2}
\end{figure}

Finally, we observe that adding two more fitting constants to construct a quadratic-log correlation for the 3D multiphysics model only reduces the mean error from 11.6\% to 10.0\%; however, the 6-parameter fit significantly improves the agreement between the medians of the correlation and the underlying physical model at the lower bound of the lubricant volume range, while sacrificing some agreement between the medians at the center of the lubricant volume range. Comparing the 3D model correlations, although the trends are similar, the quadratic-log fit clearly shows more nonlinearity with respect to lubricant volume in Fig. \ref{fig:3DCondRatioVol2} compared to the relatively linear fit in Fig. \ref{fig:3DCondRatioVol}, and it attains a slightly higher conductance ratio at the minimum load for the largest lubricant volume in Fig. \ref{fig:3DCondRatioLoad2} compared to that of the 4-parameter fit in Fig. \ref{fig:3DCondRatioLoad}.

\begin{align}\label{eq:3DModelCorrelationResult2}
    \begin{split}
    I_\text{fit, 3D, 6}\ps*{V_l;\,r_b, a_\text{3D}} & = \ps*{2.80\times 10^{-2}\cdot \ln{\frac{a_\text{3D}}{r_b}} + 1.42\times 10^{-1}}\ln^2\ps*{\frac{V_l}{\frac{4}{3} \pi r_b^3}}\\
    & + \ps*{6.27\times 10^{-1}\cdot \ln{\frac{a_\text{3D}}{r_b}} + 3.21}\ln\ps*{\frac{V_l}{\frac{4}{3} \pi r_b^3}} \\
    & + 4.73 \cdot \ln{\frac{a_\text{3D}}{r_b}} + 25.4
    \end{split}
\end{align}

\begin{figure}[H]
\centering
\begin{subfigure}{.49\textwidth}
    \centering
    \includegraphics[width=1\linewidth]{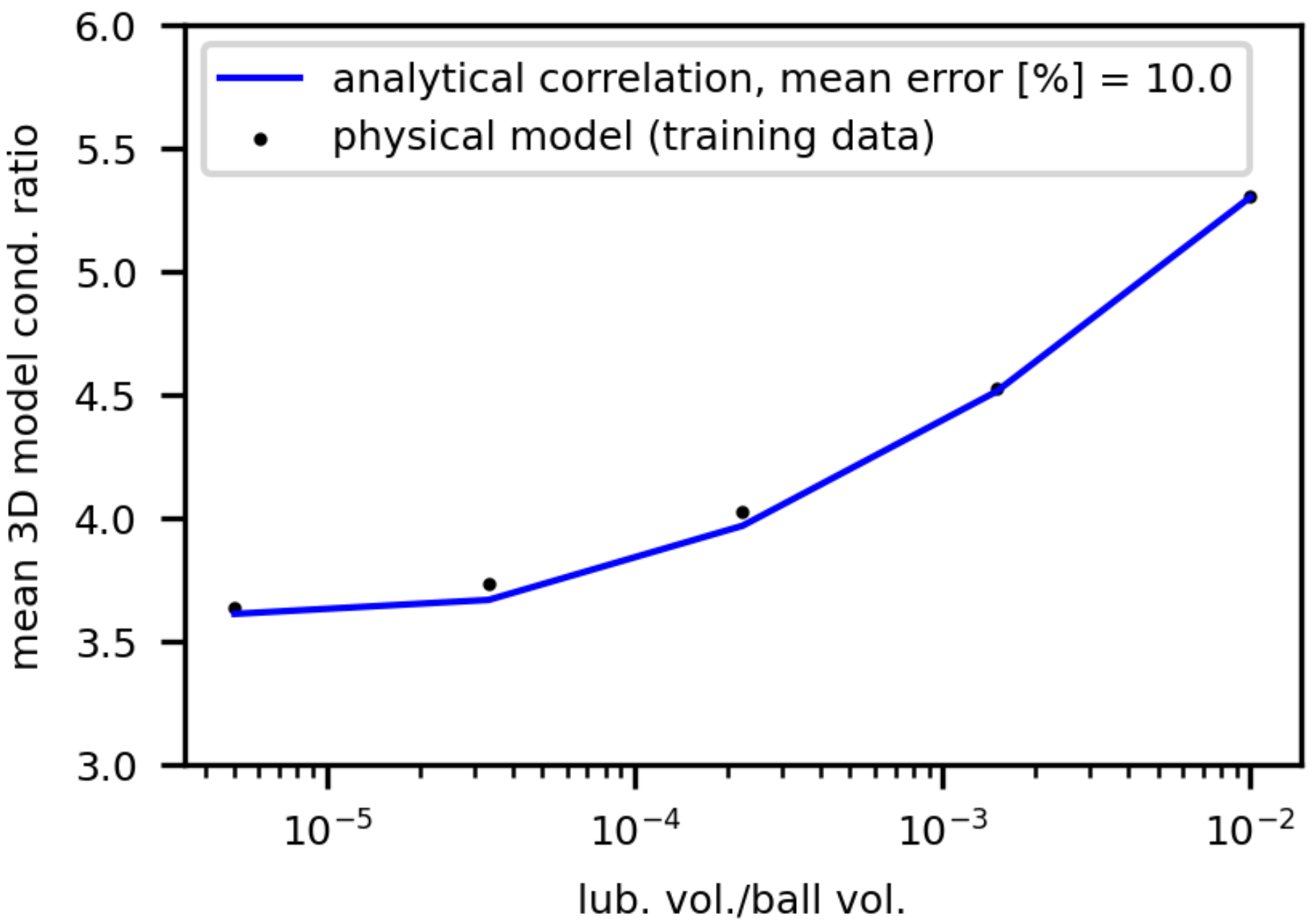}
    \caption{means, averaged across all \(r_b,\,P,\,E_r\)}
    \label{fig:3DMeans2}
\end{subfigure}
\hfill
\begin{subfigure}{.49\textwidth}
    \centering
    \includegraphics[width=1\linewidth]{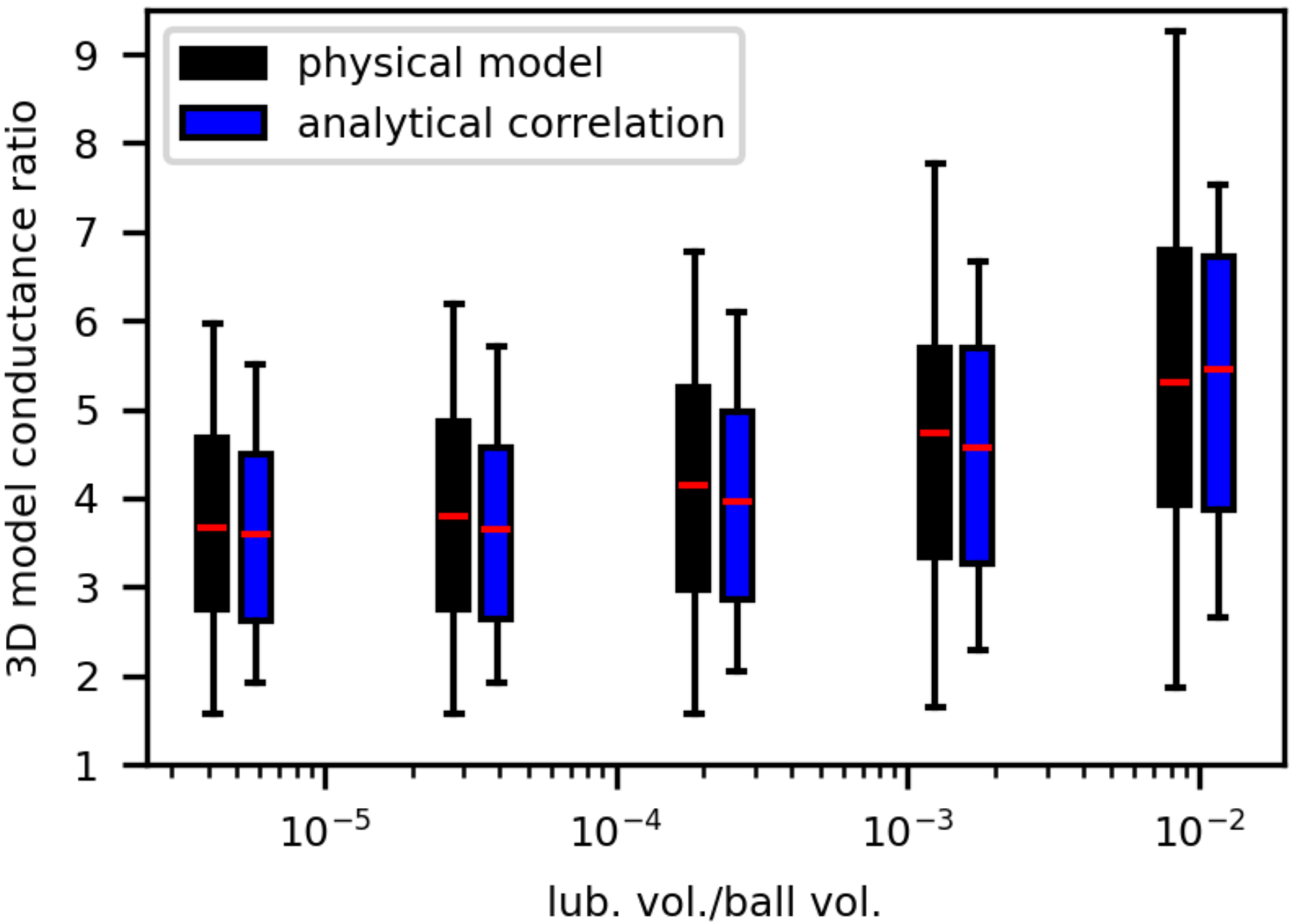}
    \caption{underlying distribution in \(r_b,\,P,\,E_r\)}
    \label{fig:3DMedians2}
\end{subfigure}
\caption{Comparison of 3D multiphysics-based analytical correlation with 6 fitting constants and physical model calculations as function of normalized lubricant volume}
\label{fig:3DMeansMedians2}
\end{figure}

\section{Conclusion}
This work demonstrates thermal conductance models for both a lubricated ball-on-flat and a lubricated angular contact ball bearing section, founded on a multiphysics approach of calculation of a lubricant meniscus combined with solution of the heat diffusion equation. By providing robust correlations to the thermal community, we enable parametric study of thermal conductance that can help theorists and experimentalists alike in elucidating effects of relevant input parameter regimes on anticipated conductances for their application. Additionally, experimentalists may find the correlations particularly useful for extrapolating the conductance of an angular contact ball bearing interface, based on a measurement taken in a physical ball-on-flat setup and comparison of the numerical predictions from the 2D and 3D multiphysics correlations detailed here. However, the underlying hierarchical methodology presented is perhaps a greater contribution than the reported correlation fits, as the methods and forms can be utilized to efficiently generate predictive correlations for any custom 2D or 3D ball-on-surface system geometries.

In the future, we would like to expand the correlations' mathematical structure to account for parameters so far kept constant, including the distance from bearing ring center to ball center \(r_\text{bearing}\), raceway radius \(r_r\), bearing contact angle \(\varphi_b\), and lubricant surface tension \(\gamma_\text{LV}\). Also, we aim to eventually support wider ranges for the input variables of ball radius \(r_b\), applied load \(P\), and reduced elastic modulus \(E_r\); increasing the number of data points underlying the correlations is key to achieving this expansion. Furthermore, it may be valuable to examine if the polynomial forms presented here are best for the 2D and 3D multiphysics models' correlation structures, as exploring other functional relationships could decrease the mean errors of the 2D and 3D multiphysics correlations beyond what we have achieved thus far. Lastly, we may be able to choose more relevant constants for the lubricant's surface tension and surface interaction strength, and in turn better match experimental thermal conductance data, by measuring lubricant contact angles in real ball-on-flat test setups.

\section{Acknowledgements}
The author gratefully acknowledge support from The Aerospace Corporation, and thank them for many fruitful discussions.

\end{document}